\documentclass[aps,twocolumn,preprintnumbers,showpacs,superscriptaddress,nofootinbib,showpacs,amsmath,groupedaddress,floatfix,secnumarabic]{revtex4}

\usepackage[english]{babel}
\usepackage[utf8]{inputenc}

\usepackage[normalem]{ulem}

\usepackage{epsfig,verbatim,amssymb,amsfonts,graphicx}
\usepackage{subfigure}

\usepackage[colorlinks]{hyperref}

\hypersetup{colorlinks=true,citecolor=blue,linkcolor=blue,urlcolor=blue}

\usepackage{xcolor}

\usepackage{color}

\def\scn#1#2{\section{#1}\lb{#2}}

\def\bfl{\begin{flushleft}}
\def\efl{\end{flushleft}}
\def\bfr{\begin{flushright}}
\def\efr{\end{flushright}}
\def\bc{\begin{center}}
\def\ec{\end{center}}
\def\be{\begin{equation}}
\def\ee{\end{equation}}
\def\bse{\begin{subequations}}
\def\ese{\end{subequations}}
\def\ba{\begin{eqnarray}}
\def\ea{\end{eqnarray}}
\def\baa#1{\begin{array}{#1}}
\def\eaa{\end{array}}
\def\bw{\begin{widetext}}
\def\ew{\end{widetext}}

\def\lb#1{\label{#1}}
\def\bit{\begin{itemize}}
\def\eit{\end{itemize}}
\def\bco{}
\def\bcs{\begin{cases}}
\def\ecs{\end{cases}}

\def\nc0{\tilde b_0}

\def\phie{\vartheta}

\begin{document}

\preprint{\small Phys. Lett. B \textbf{823}, 136776 (2021)   
\ [\href{https://doi.org/10.1016/j.physletb.2021.136776}{DOI: 10.1016/j.physletb.2021.136776}]
}

\title{Kink solutions in logarithmic scalar field theory: Excitation spectra, scattering, and decay of bions}

\author{Ekaterina Belendryasova}
\email{7.95@bk.ru}
\affiliation{A.A. Bochvar High-Technology Scientific Research Institute for Inorganic Materials (VNIINM), Moscow 123098, Russia}
\affiliation{National Research Nuclear University MEPhI (Moscow Engineering Physics Institute), Moscow 115409, Russia}

\author{Vakhid A. Gani}
\email{vagani@mephi.ru}
\affiliation{National Research Nuclear University MEPhI (Moscow Engineering Physics Institute), Moscow 115409, Russia}
\affiliation{Institute for Theoretical and Experimental Physics of National Research Centre ``Kurchatov Institute'', Moscow 117218, Russia}

\author{Konstantin G. Zloshchastiev}
\email{https://bit.do/kgz}
\affiliation{Institute of Systems Science, Durban University of Technology, P.O. Box 1334, Durban 4000, South Africa}



\begin{abstract}
We consider the (1+1)-dimensional Lorentz-symmetric field-theoretic model with logarithmic potential having a Mexican-hat form with two local minima similar to that of the quartic Higgs potential in conventional electroweak theory with spontaneous symmetry breaking and mass generation. We demonstrate that this model allows topological solutions --- kinks. We analyze the kink excitation spectrum, and show that it does not contain any vibrational modes. We also study the scattering dynamics of kinks for a wide range of initial velocities. The critical value of the initial velocity occurs in kink-antikink collisions, which thus differentiates two regimes. Below this value, we observe the capture of kinks and their fast annihilation; while above this value, the kinks bounce off and escape to spatial infinities. Numerical studies show no resonance phenomena in the kink-antikink scattering.
\end{abstract}



\maketitle

\section{Introduction}

Since the works \cite{ros68,ros69}, relativistic scalar field theories with logarithmic nonlinearity have 
attracted significant interest. 
Such models 
were proved to be  instrumental in the theory 
of quantum fields and particles
\cite{ros68,ros69,em98,dz12},
quantum liquids and superfluidity \cite{az11,z18zna,z19ijmpb,sz19}, 
fluid mechanics \cite{z18zna,z18epl,lz21cs},
and the
theory of physical vacuum, classical and quantum gravity \cite{z10gc,z11appb,szm16,z20ijmpa,z20un1,z21ltp}.

The mathematical properties of solutions of logarithmic wave equations, 
both in the Lorentz-symmetric and Galilean cases, 
have been extensively studied since the works \cite{ros68,bbm75}. 
One of the model's most striking features is the occurrence of localized solutions, which have a Gaussian profile.
For this reason they are often referred to as gaussons in non-relativistic literature. 
Their relativistic analogues have proven useful for describing extended particles and $Q$-balls in high energy particle physics \cite{ros68,dz12}. 

Naturally, one can ask a question: 
whether the logarithmic scalar field theories of the above-mentioned type have 
topologically nontrivial solutions, such as {\it kinks}. 
This question is not only of mathematical interest, 
but also has a direct relevance to physics: 
even in the (1+1)-dimensional case, kink solutions find their application in quantum and classical field theory, high energy physics, cosmology, and condensed matter theory; 
further details can be found, e.g., in Ref.~\cite{Vilenkin.book.2000}.

The important feature of the (1+1)-dimensional models with one field is 
that they can be studied, both analytically and numerically, far more easily 
than the (2+1)- or (3+1)-dimensional models with multiple fields.
Moreover, many physical systems can be approximately or effectively described by spatially one-dimensional one-field configurations.
For example, a plane domain wall separating regions with different vacuum states can be viewed as a kink configuration along the direction normal to the wall.

Kink-type solutions occur in a number of models, with both polynomial and non-polynomial self-interaction of a field. Important results have been obtained for models with polynomial potentials of the fourth and higher degrees \cite{Kevrekidis.book.2019,Dorey.PRL.2011,Gani.PRD.2014,Moradi.JHEP.2017,Khare.PRE.2014,Gani.JHEP.2015,Belendryasova.CNSNS.2019,Christov.PRD.2019,Christov.PRL.2019,Manton.JPA.2019,Khare.JPA.2019,Gani.PRD.2020,Christov.CNSNS.2021,Campos.arXiv.2020}.
As an example of non-polynomial models, we can mention modifications of the sine-Gordon model,
see, e.g., Refs.~\cite{Campbell.PhysD.1986.dsG,Gani.EPJC.2018,Gani.EPJC.2019}.
It is worth mentioning the so-called deformation procedure, which allows one to find a great variety of models and their kink-type solutions \cite{Bazeia.PRD.2002,Blinov.arXiv.2020.deform,Blinov.JPCS.2020.deform}.
Another recent direction of study is long-range interactions between kinks and antikinks \cite{Belendryasova.CNSNS.2019,Christov.PRD.2019,Christov.PRL.2019,Manton.JPA.2019,Khare.JPA.2019,Christov.CNSNS.2021,Campos.arXiv.2020,Guerrero.PLA.1998,Gomes.PRD.2012}. 
Substantial research has also been done for non-minimal point particles and zero-branes \cite{zlo00mpla,zlo01plb}, domain walls, bubbles \cite{Gani.JHEP.2016,Campanelli.IJMPD.2004,Gonzalez.JCAP.2018}, $Q$-balls
\cite{dz12,Nugaev.PRD.2013,Bazeia.EPJC.2016},
and various phenomena occurred in the early Universe \cite{Gani.JCAP.2018.earlyUni}.

In the non-relativistic case, kink solutions were explicitly obtained in Ref.~\cite{z18epl} and further studied in Ref.~\cite{z19ijmpb}.
Preliminary studies of kink solutions in relativistic logarithmic nonlinear scalar field theory were reported in Refs.~\cite{z11appb,Belendryasova.JPCS.2019.logKinks}.
In this paper, we study the spectrum of small excitations and the scattering properties of relativistic kink-type solutions.

In general, kink scattering phenomena, as well as the interactions of kinks with impurities (spatial inhomogeneities), 
have been of growing interest since the 70's \cite{Belova.UFN.1997}; 
and are still a rapidly developing area of research. 
In this connection, among analytical methods, one can mention the collective coordinate approach \cite{Gani.PRD.2014,Christov.PRD.2019,Belova.UFN.1997,Weigel.PRD.2016.cc,Demirkaya.JHEP.2107.cc,Baron.JPhysA.2014}, where a kink-antikink field configuration is approximately described as a system with one or several degrees of freedom.
On the other hand, numerical methods have recently become a powerful tool for studying the dynamics of various field phenomena. 
In particular, resonance phenomena, such as escape windows and quasiresonances, have been discovered in kink scattering \cite{Dorey.PRL.2011,Gani.PRD.2014,Gani.JHEP.2015,Moradi.JHEP.2017,Belendryasova.CNSNS.2019,Campbell.PhysD.1986.dsG,Gani.EPJC.2018,Gani.EPJC.2019,Christov.CNSNS.2021,Bazeia.EPJC.2018.sinh}. 

The paper is organized as follows. 
In Sec.~\ref{s:field}, we formulate the relativistic scalar field theory with logarithmic nonlinearity.
We show that this theory allows topologically nontrivial solutions --- kinks. 
In Sec. \ref{s:kexc}, we focus on the kink solution, perform its linear stability analysis, and study the scattering properties of the kink-antikink configuration.
Then, in Sec.~\ref{s:higgs}, we give a brief comparative discussion of the logarithmic, the $\phi^4$, and some other field models. Finally, we summarize and conclude in Sec.~\ref{s:con}.

\scn{Logarithmic scalar model and its kink solution}{s:field}

Let us consider a (1+1)-dimensional field-theoretic model with a real scalar field $\phi(x,t)$.
Our working model is defined by the Lagrangian density
\be\label{eq:Largangian}
\mathcal{L} = 
\frac{1}{2}\left(\frac{\partial\phi}{\partial t}\right)^2 - \frac{1}{2}\left(\frac{\partial\phi}{\partial x}\right)^2 - V(\phi),
\ee
with the following potential term:
\be\lb{e:pot}
V (\phi) = - b\, \phi^2 \left[ \ln{(\phi^2/\phie^2) - 1} \right] + V_0^{}
,
\ee
where $b$, $\phie$, and $V_0^{}$ are real-valued constant parameters, $\phie>0$.

\begin{figure}[t!]
\centering\includegraphics[width=0.7\linewidth]{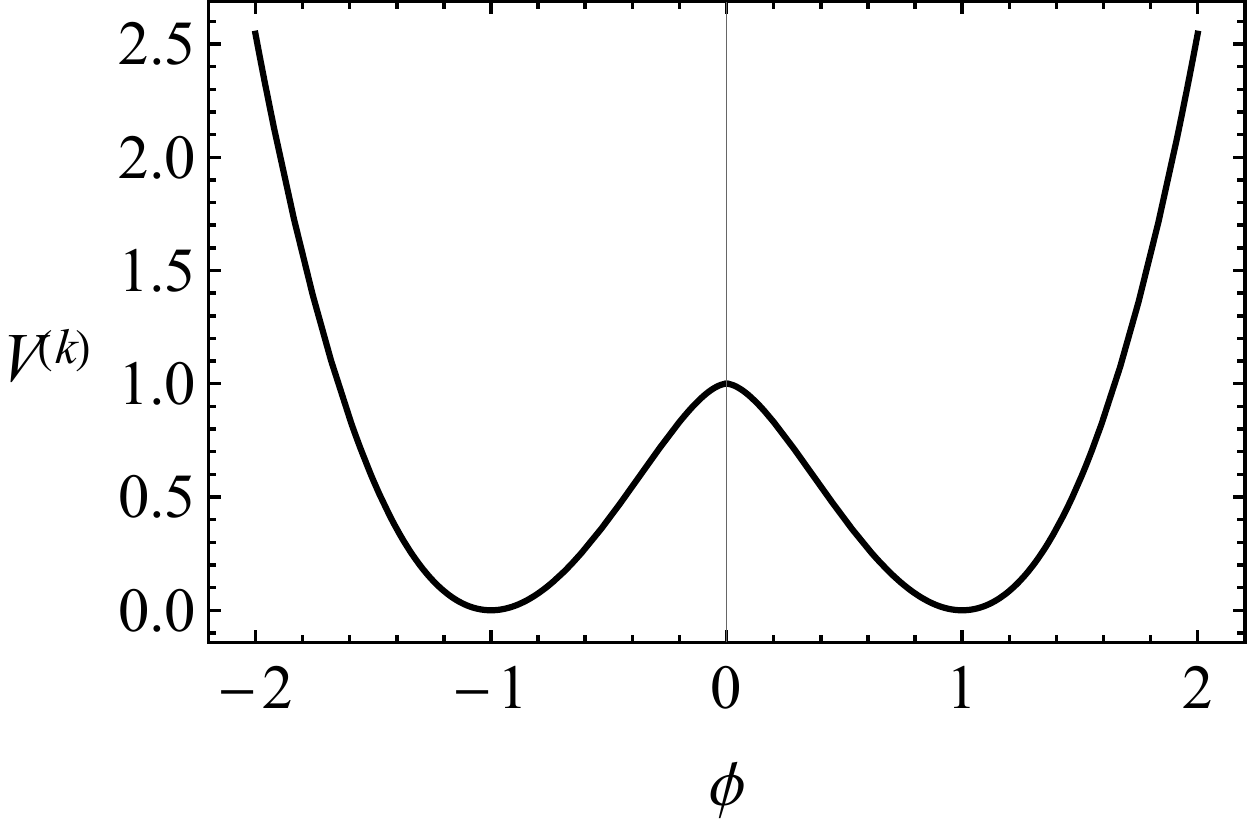}
\caption{Potential \eqref{e:potk} in units of $\vartheta^2/\ell^{2}$, versus $\phi$ in units $\vartheta$.}
\label{f:fpotkin}
\end{figure}

Let us consider a model for which 
$b<0$ and
$V_0^{}=-b \vartheta^2 = |b| \vartheta^2$.
Then the potential \eqref{e:pot} becomes
\be\lb{e:potk}
V^{\rm (k)}(\phi)=\frac{\phi^2}{\ell^2} \left[\ln{(\phi^2/\phie^2) - 1}\right] + \frac{\phie^2}{\ell^2},
\ee
where $\ell = 1/\sqrt{-b}$ is a real-valued parameter.
The potential \eqref{e:potk} has a Mexican-hat form, with local minima at $|\phi| = \phie$, see Fig.~\ref{f:fpotkin}.
Thus, our model is somewhat similar to the quartic, or the $\phi^4$, model, widely used for describing systems with spontaneous symmetry breaking.
In fact, the $\phi^4$ model can be regarded as one of the perturbative limits of the logarithmic model \cite{az11}.
For instance, if one expands potential \eqref{e:potk} in the Taylor series near its local minima then one obtains
\be\lb{e:potkapp}
V^{\rm (k)}(\phi)
\approx
\frac{\lambda_{\rm H}^{}}{4}
\left(\phi^2-\phie^2\right)^2
,
\ee
where $\lambda_{\rm H}^{} = 2/(\phie^2 \ell^2)$.
Notice that the coefficient of the quadratic term, $- 1/\ell^2$, is negative, thus it cannot be regarded as a squared mass of a scalar particle.
Therefore, the logarithmic kink model describes, not a point-particle system, but a collective non-local state, such as the quantum liquid.

The kink solution of the model \eqref{e:potk} satisfies the first order differential equation:
\be\lb{e:ssgenk}
\frac{d\phi}{dx} = \pm \frac{\sqrt{2}}{\ell}\left|\phi\right|\sqrt{\ln{(\phi^2/\phie^2)-1} + \frac{\phie^2}{\phi^2}}
,
\ee
which should be integrated with taking into account additional constraint fixing the kink position on the $x$-axis.

\begin{figure}[t!]
\subfigure[\ scalar field, in units $\vartheta$]{
\centering\includegraphics[width=0.7\linewidth]{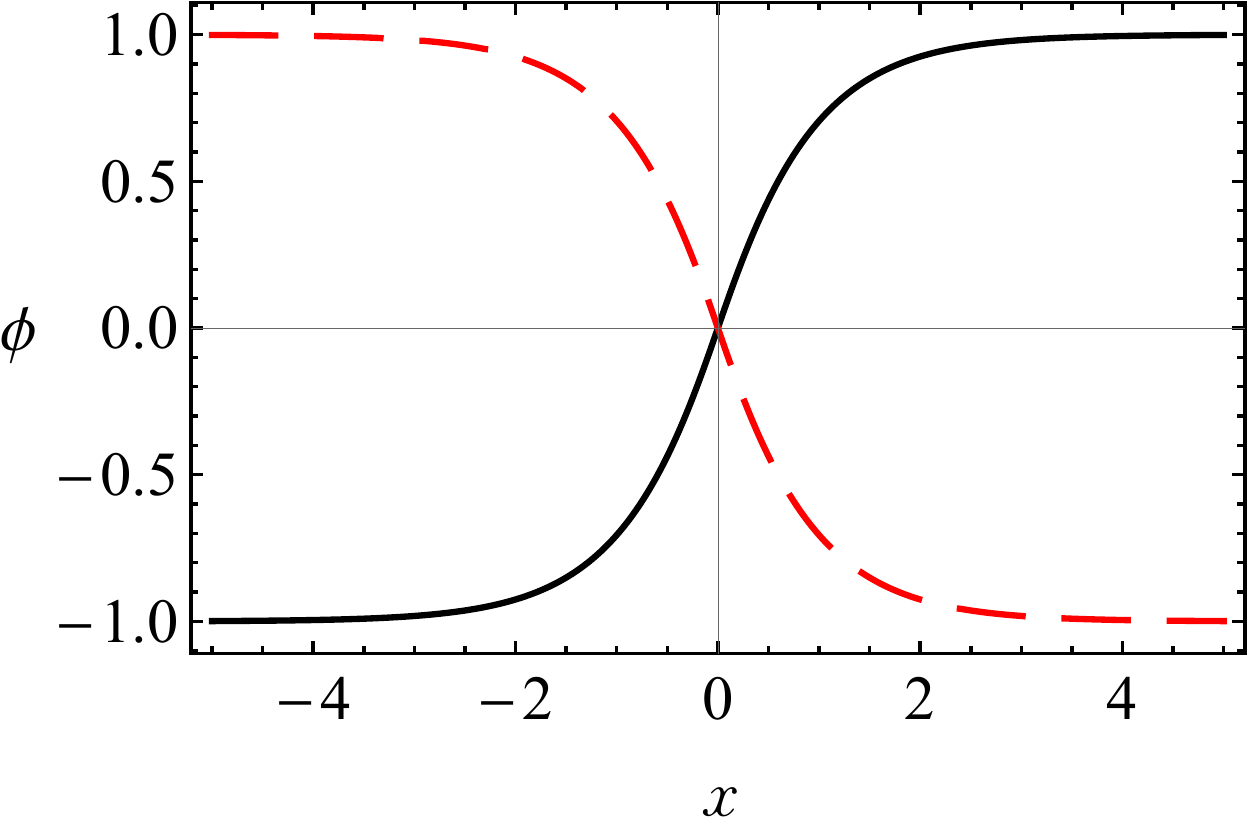}
}
\subfigure[\ energy density, in units of $\vartheta^2/\ell^2$]{
\centering\includegraphics[width=0.7\linewidth]{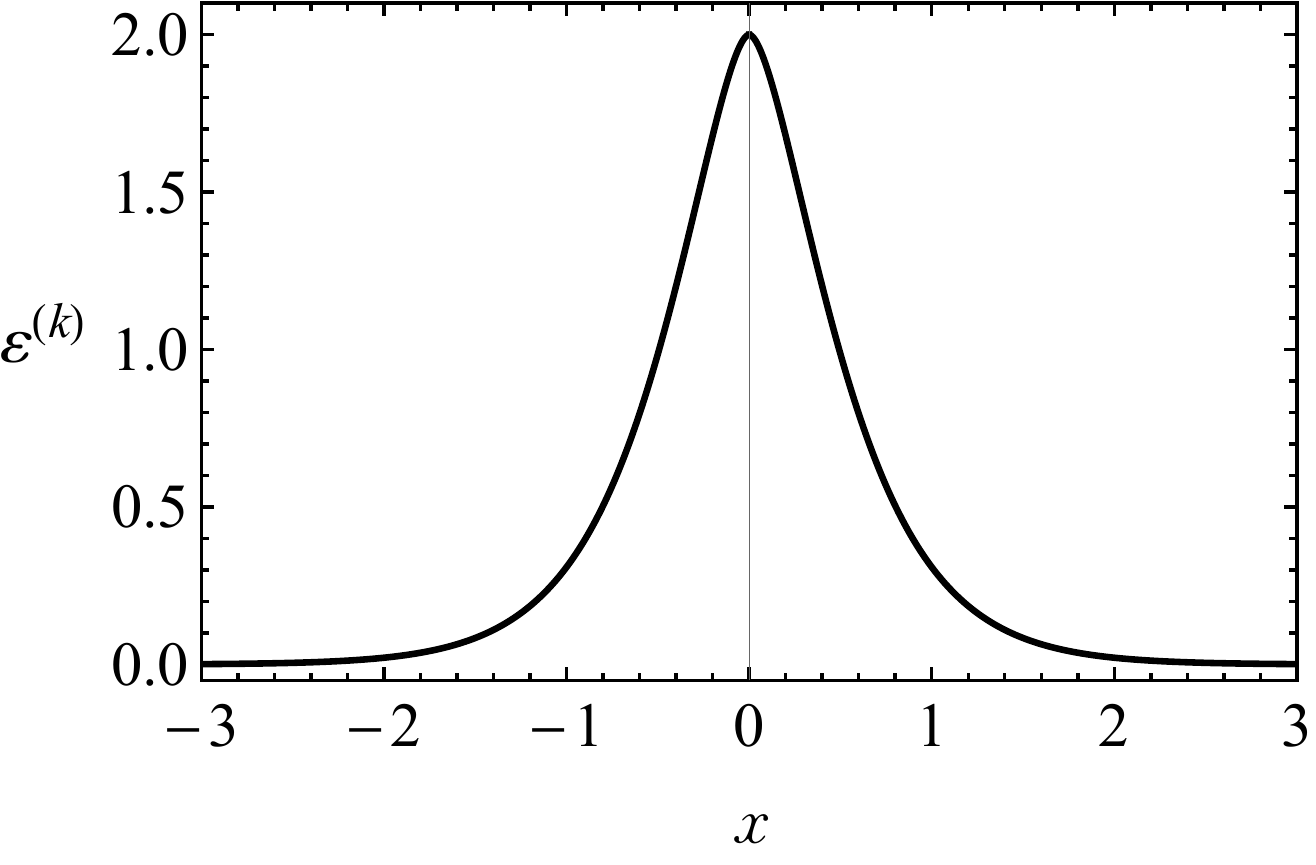}
}
\caption{Profiles of the kink 
(solid curve) and antikink 
(dashed curve) solutions, 
and their energy densities, versus $x/\ell$.
}\label{fig:kinks}
\end{figure}

An exact analytical solution of Eq.~\eqref{e:ssgenk} is unknown, therefore, we have to resort to numerical computations.
In doing so, we obtain the {\it kink} solution, which is a monotonous function interpolating between the nontrivial vacua of the model,
while the {\it antikink} solution can be obtained by the mirror reflection with respect to the vertical axis, see Fig.~\ref{fig:kinks}(a).

The kink and antikink solutions have the same energy densities $\varepsilon^\mathrm{(k)}$, see Fig.~\ref{fig:kinks}(b),
and their total energy can be computed as
\be
E^\mathrm{(k)}_\pm = \frac{\sqrt{2} }{\ell} \vartheta^2 \int\limits_{-1}^{1} \sqrt{s^2(\ln{s^2}-1)+1}\ ds
\approx 
1.187 \frac{\sqrt{2}}{\ell} \vartheta^2    
,
\ee
where we used a rescaling $s=\phi/\vartheta$ and calculated the integral numerically.

Note that $l$ and $\phie$ can be eliminated by the appropriate choice of units for coordinates and field. This is equivalent to the fact that $l$ and $\phie$ can be set equal to any fixed constants. We will use this option below.

In the following sections of the paper, we will focus our attention on the properties of the kink solutions.

\scn{Kink excitation spectrum and scattering of kinks}{s:kexc}

Here let us present an analysis of the linear stability of the kink solution. 
To compute the kink excitation spectrum, we add a small perturbation $\delta\phi(x,t)$ to the static kink solution $\phi^{\rm (k)}(x)$:
\begin{equation}
\phi(x,t) = \phi^{\rm (k)}(x) + \delta\phi(x,t), \quad ||\delta\phi|| \ll ||\phi^{\rm (k)}||
,
\end{equation}
and plug it into the equation of motion. Taking into account only linear terms with respect to $\delta\phi$, we obtain the following equation
\begin{equation}\label{eq:delta_phi}
\frac{\partial^2\delta\phi}{\partial t^2} - \frac{\partial^2\delta\phi}{\partial x^2} + \left.\frac{d^2 V^{\rm (k)}}{d\phi^2}\right|_{\phi^{\rm (k)}(x)}\cdot\delta\phi = 0
,
\end{equation}
whose solution can be found using separation of variables $x$ and $t$:
\begin{equation}
\delta\phi(x,t) = \eta(x)\cos\:\omega t.
\end{equation}
Then we obtain from Eq.~\eqref{eq:delta_phi}
the eigenvalue problem --- the time-independent Schr\"odinger-like equation:
\begin{equation}\label{eq:stat_Schr}
\hat{H}\eta(x) = \omega^2\eta(x),
\end{equation}
where the Hamiltonian operator is
\begin{equation}\label{eq:Schr_Ham}
\hat{H} = -\frac{d^2}{dx^2} + U(x)
\end{equation}
with
\begin{equation}\label{eq:Schr_pot}
U(x) = \left.\frac{d^2 V^{\rm (k)}}{d\phi^2}\right|_{\phi^{\rm (k)}(x)}
\end{equation}
being the {\it stability potential}, sometimes referred to as the \textit{quantum-mechanical potential}.

The excitation spectra of the kink and antikink solutions are obviously the same, 
therefore, let us consider the kink interpolating between vacua $\phi=-\phie$ and $\phie$.
\begin{figure}[t!]
\centering\includegraphics[width=0.7\linewidth]{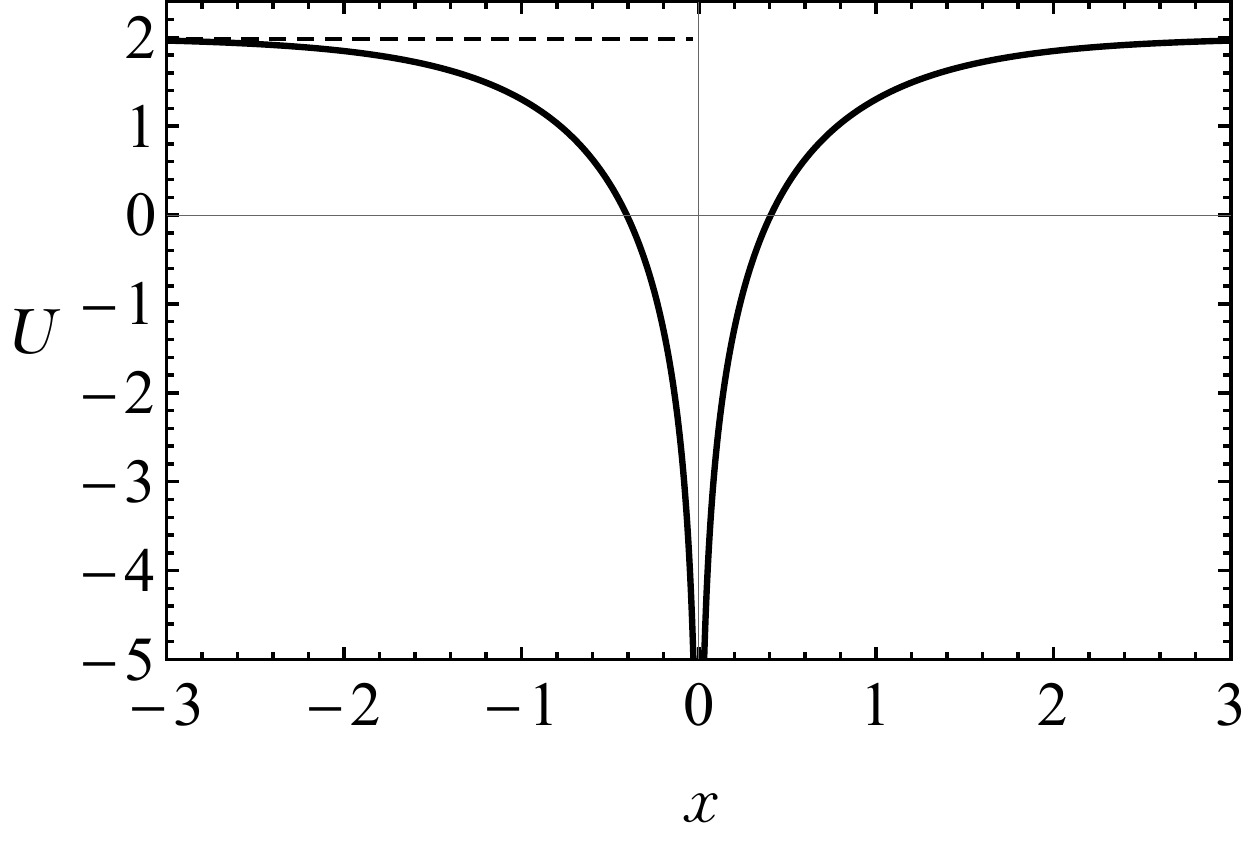}
\caption{Profile of the stability potential 
\eqref{eq:Schr_pot_k} versus $x$; $\ell=\sqrt{2}$, $\phie=1$.}
\label{fig:Schr_pot}
\end{figure}
The corresponding stability potential \eqref{eq:Schr_pot} is
\begin{equation}\lb{eq:Schr_pot_k}
U(x) = \frac{2}{\ell^2}
\left\{
\ln\left[
\frac{\left(\phi^{\rm (k)}(x)\right)^2}{\phie^2}
\right]
+2 
\right\}
,
\end{equation}
see Fig.~\ref{fig:Schr_pot}.
At $x\to 0$, we obtain $\phi^{\rm (k)}(x)\to 0$ and $U(x)\sim\ln x^2\to -\infty$.

Furthermore, it can easily be shown, that no negative values of $\omega^2$ exist in the kink excitation spectrum. Moreover, the function ${d\phi^{\rm (k)}}/{dx}$ is the eigenfunction corresponding to the eigenvalue $\omega_0^{}=0$, i.e.\ kink always has a zero (translational) mode, see, e.g., Sec.~2 in \cite{Gani.JHEP.2015}.

\begin{figure}[t!]
\centering\includegraphics[width=0.7\linewidth]{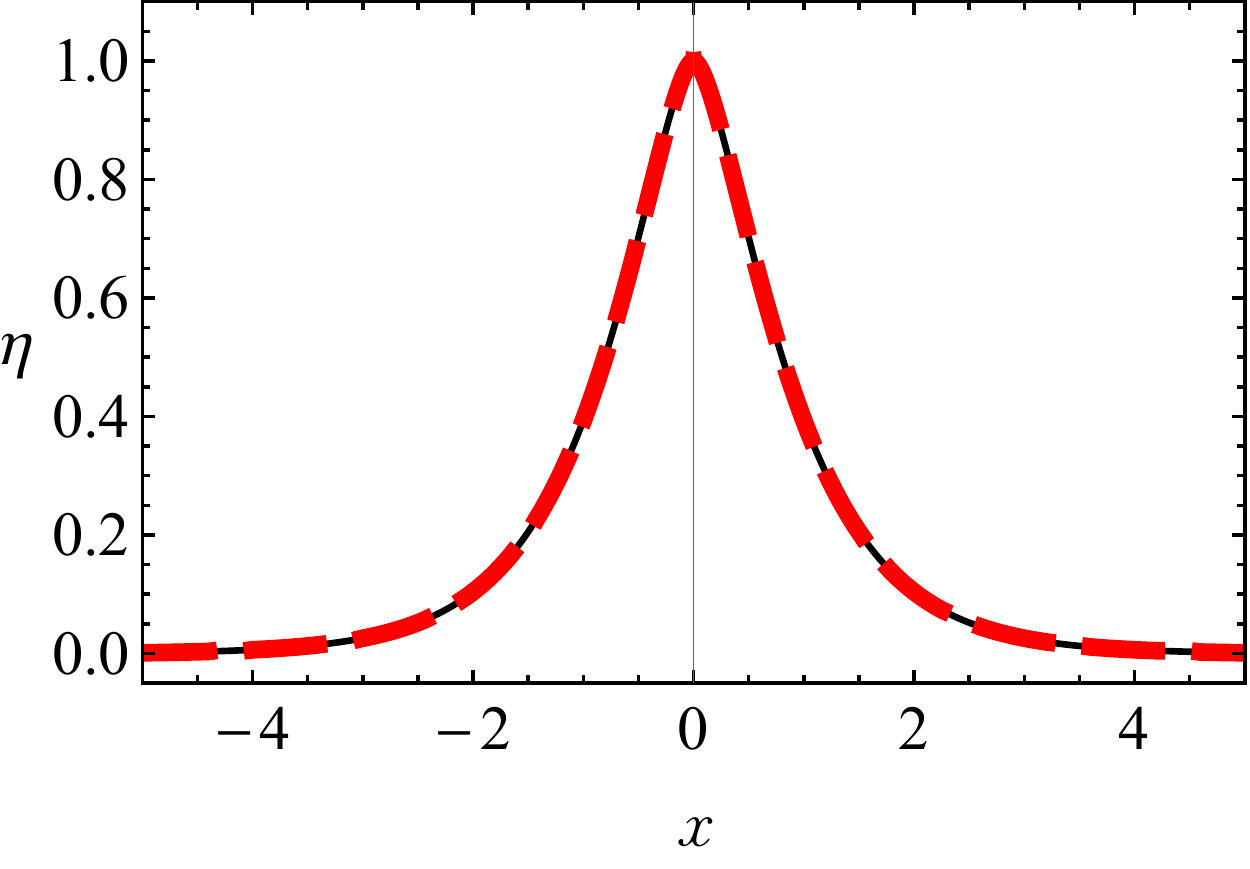}
\caption{Profiles of the eigenfunction of zero (translational) mode $d\phi^{\rm (k)}/dx$ (black solid curve), and (appropriately normalized) eigenfunction of the problem \eqref{eq:stat_Schr} corresponding to the eigenvalue $\omega_0^{}$ (red dashed curve) versus $x$; $\ell=\sqrt{2}$, $\phie=1$.
}
\label{fig:zero_mode_eigenfunction}
\end{figure}

To search for eigenstates, we used the finite element method \cite{fem-segerlind},
in which eigenvalues and eigenvectors are calculated using Schur decomposition.
We used uneven discretization with a varying step, with a minimal value of 0.005. 
We found the only frequency $\omega_0^{}\approx 0.07$ in the descrete spectrum,
its corresponding eigenfunction is shown in Fig.~\ref{fig:zero_mode_eigenfunction}.
The absence of vibrational mode(s) indicates that probably no resonance phenomena occur in kink-antikink scattering. Nevertheless, experimental studies of kink collisions are of great interest, because the appearance of resonance phenomena is possible even in the absence of vibrational modes. In the previous literature, two possibilities are known for resonant energy exchange in absence of vibrational modes in the kink's excitation spectrum:

(i) In the case of asymmetric kink and antikink, the transfer of kinetic energy to the vibrational mode of the composite kink-antikink system is possible, which was demonstrated for the $\phi^6$ kinks in Ref.~\cite{Dorey.PRL.2011}. The same idea was applied to explain escape windows in the collisions of asymmetric kinks and antikinks of the $\phi^8$ model \cite{Belendryasova.CNSNS.2019}. One could also mention the recent paper \cite{Bazeia.EPJC.2021} discussing the composite potential of the kink and antikink.

(ii) Resonant exchange of kinetic energy with the quasinormal mode can occur, see, e.g., Ref.~\cite{Dorey.PLB.2018}. If the lifetime of the quasinormal mode is large enough, it can accumulate part of kinetic energy after the first collision of kinks. Subsequently, if the resonance condition is fulfilled, in the next collision, the energy returns to the kinetic part and kinks are able to escape to spatial infinity. (The above scenario corresponds to the two-bounce window.)

In our case, the stability potential \eqref{eq:Schr_pot_k} is symmetric, which means that the first option does not take place. As for the presence of quasinormal modes in the potential \eqref{eq:Schr_pot_k}, they probably do not exist due to the shape of the potential. Nevertheless, without a detailed study of the continuous spectrum and its features, one cannot exclude the possibility of some resonant phenomena involving frequencies from the continuous spectrum. Therefore, the numerical study of kink-antikink collisions is important.

\begin{figure}[t!]
\centering\includegraphics[width=0.7\linewidth
]{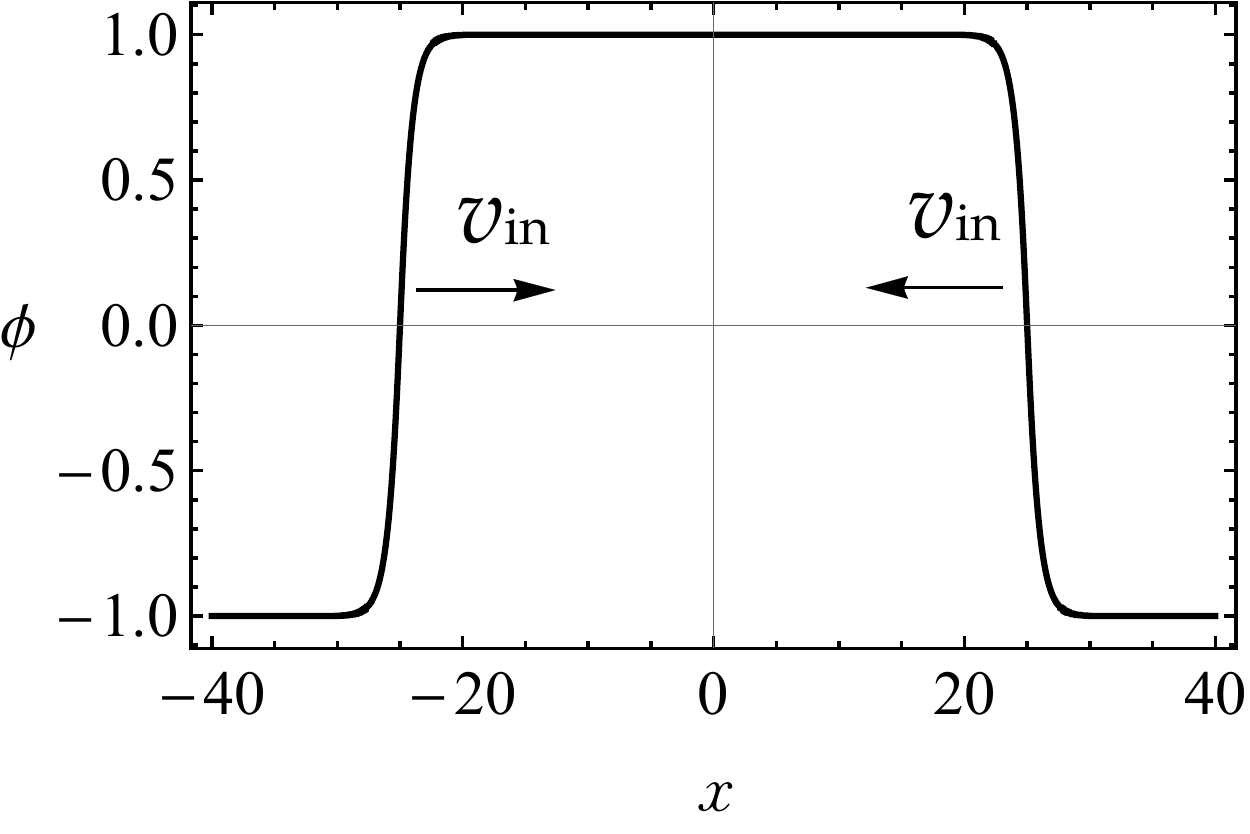}
\caption{Initial configuration for the kink-antikink scattering: scalar field $\phi$ versus $x$; $\ell=\sqrt{2}$, $\phie=1$.
}
\label{fig:init_cond}
\end{figure}

Now let us study the scattering of the kink solutions of the model.
We assume the initial configuration to be in the form of kink and antikink centered at $x=-\xi$ and $x=\xi$, respectively, and moving towards each other with initial velocities $v_{\rm in}^{}$:
\be\label{eq:init_cond}
\resizebox{\columnwidth}{!}{
$\phi(x,t)  =  
\phi_{(-1,1)}^{}\left(\frac{x+\xi-v_{\rm in}^{}t}{\sqrt{1-v_{\rm in}^2}}\right) 
+\phi_{(1,-1)}^{}\left(\frac{x-\xi+v_{\rm in}^{}t}{\sqrt{1-v_{\rm in}^2}}\right)
-1,$}
\ee
see Fig.~\ref{fig:init_cond}.

Let us assume $\phie=1$ and $\ell=\sqrt{2}$ throughout this section. 
In Eq.~\eqref{eq:init_cond}, fields $\phi_{(-1,1)}^{}(x)$ and $\phi_{(1,-1)}^{}(x)$ describe the kink interpolating between the vacua $\phi=-1$ and $\phi=1$ and the antikink interpolating between the vacua $\phi=1$ and $\phi=-1$, respectively, i.e., $\phi_{(-1,1)}^{}(x)\equiv\phi^{\rm (k)}(x)$ and $\phi_{(1,-1)}^{}(x)\equiv-\phi^{\rm (k)}(x)$. 
All necessary data for the numerical computations can be extracted from Eq.~\eqref{eq:init_cond}. 

The evolution of the initial configuration of the type $(-1,1,-1)$ is obtained by the numerical solution of the equation of motion, using the explicit finite difference scheme with discretization of the second order with $0.001$ and $0.005$ for temporal and spatial steps, and the initial half-distance $\xi=25$.

\begin{figure*}[t!]
\subfigure[\ $v_{\rm in}=0.2000$]{
\centering\includegraphics[width=
0.35\linewidth
]{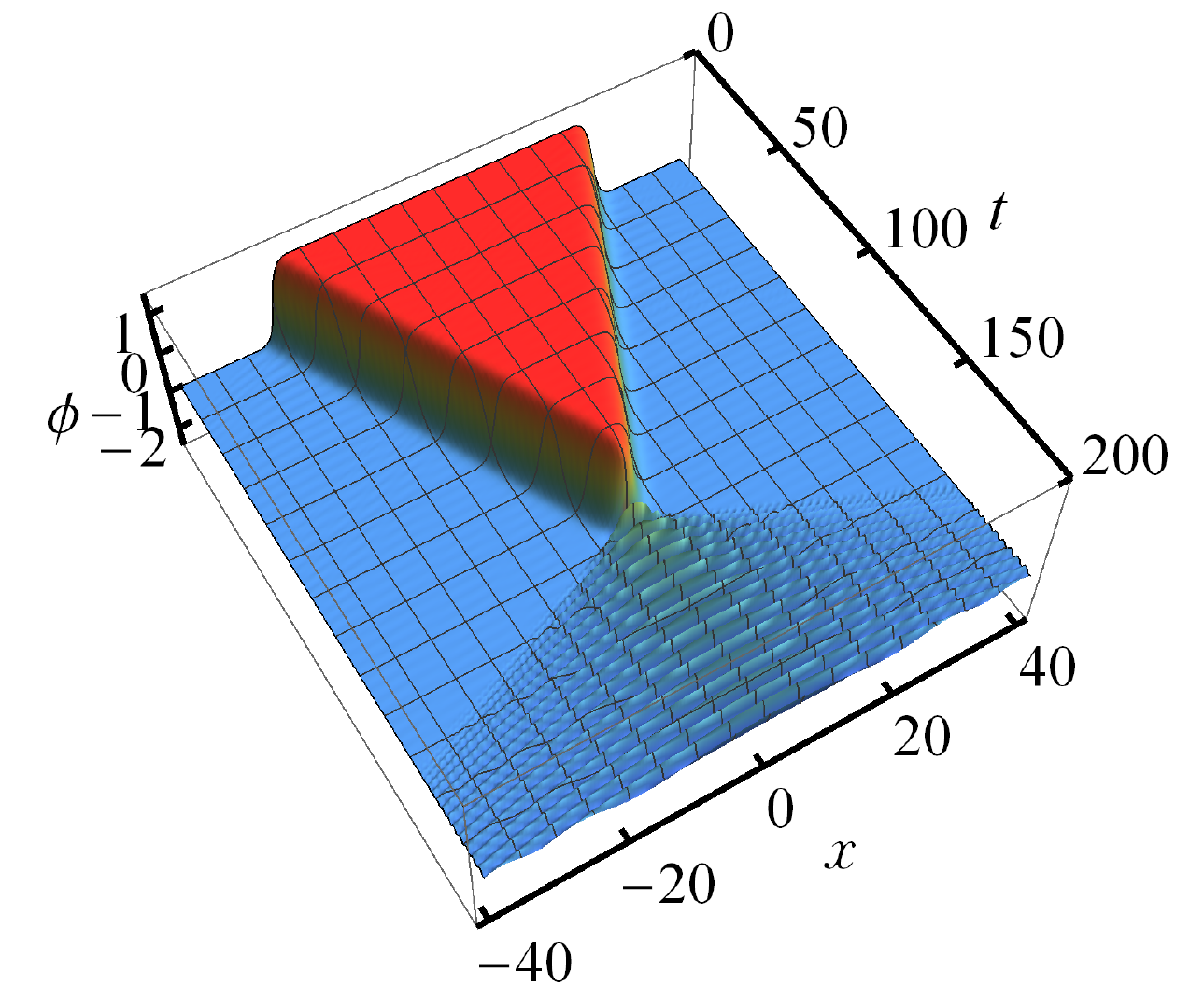}
}
\subfigure[\ $v_{\rm in}=0.5000$]{
\centering\includegraphics[width=
0.35\linewidth
]{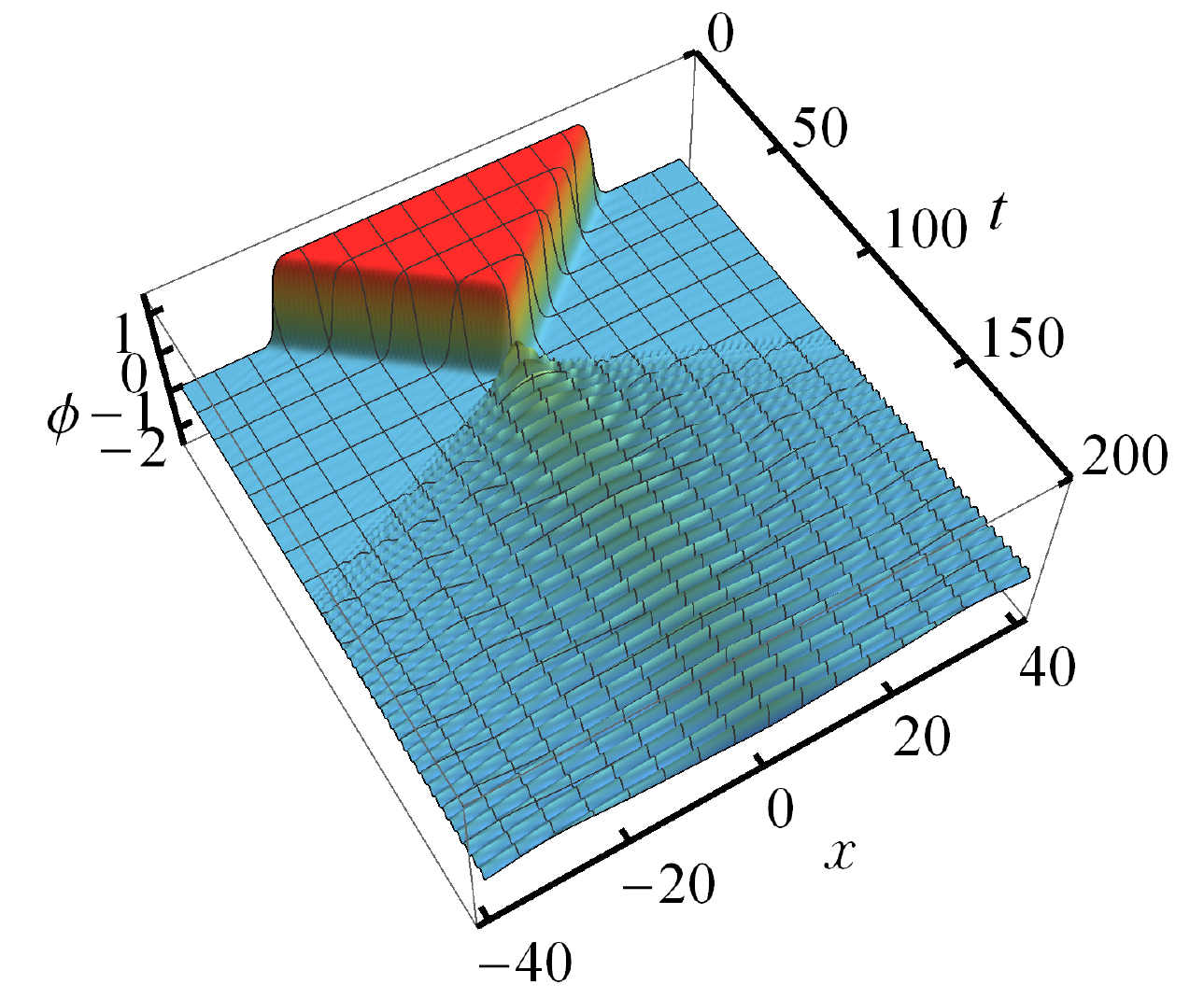}
}
\subfigure[\ $v_{\rm in}=0.7840$]{\centering\includegraphics[width=
0.35\linewidth
]{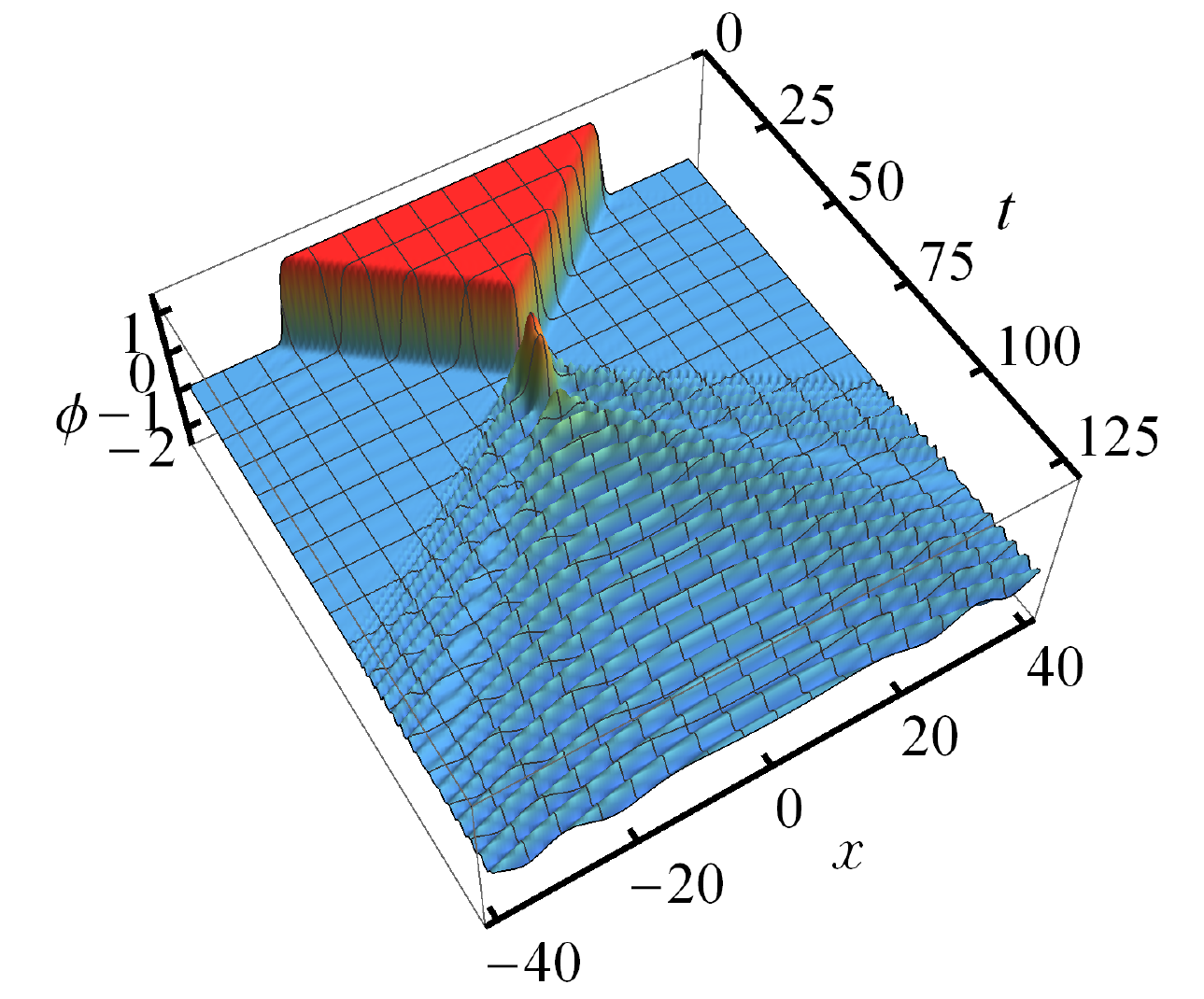}
}
\subfigure[\ $v_{\rm in}=0.7940$]{\centering\includegraphics[width=
0.35\linewidth
]{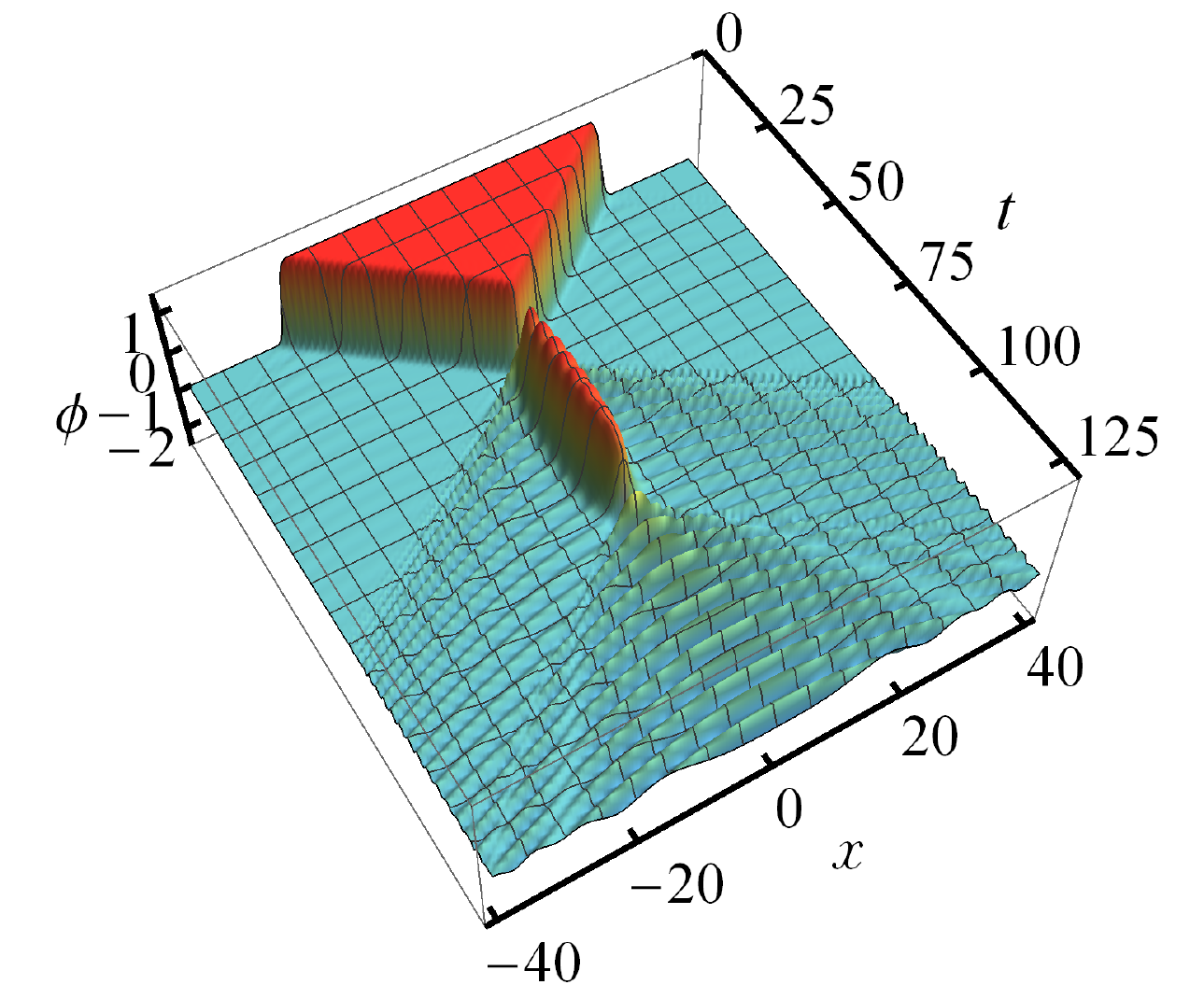}
}
\caption{Formation of a bion in kink-antikink collisions at $v_{\rm in}^{}<v_{\rm cr}^{}$; $\ell=\sqrt{2}$, $\phie=1$.
}
\label{fig:bion}
\end{figure*}

\begin{figure*}[t!]
\subfigure[\ $v_{\rm in}=0.2000$]{
\centering\includegraphics[width=
0.35\linewidth
]{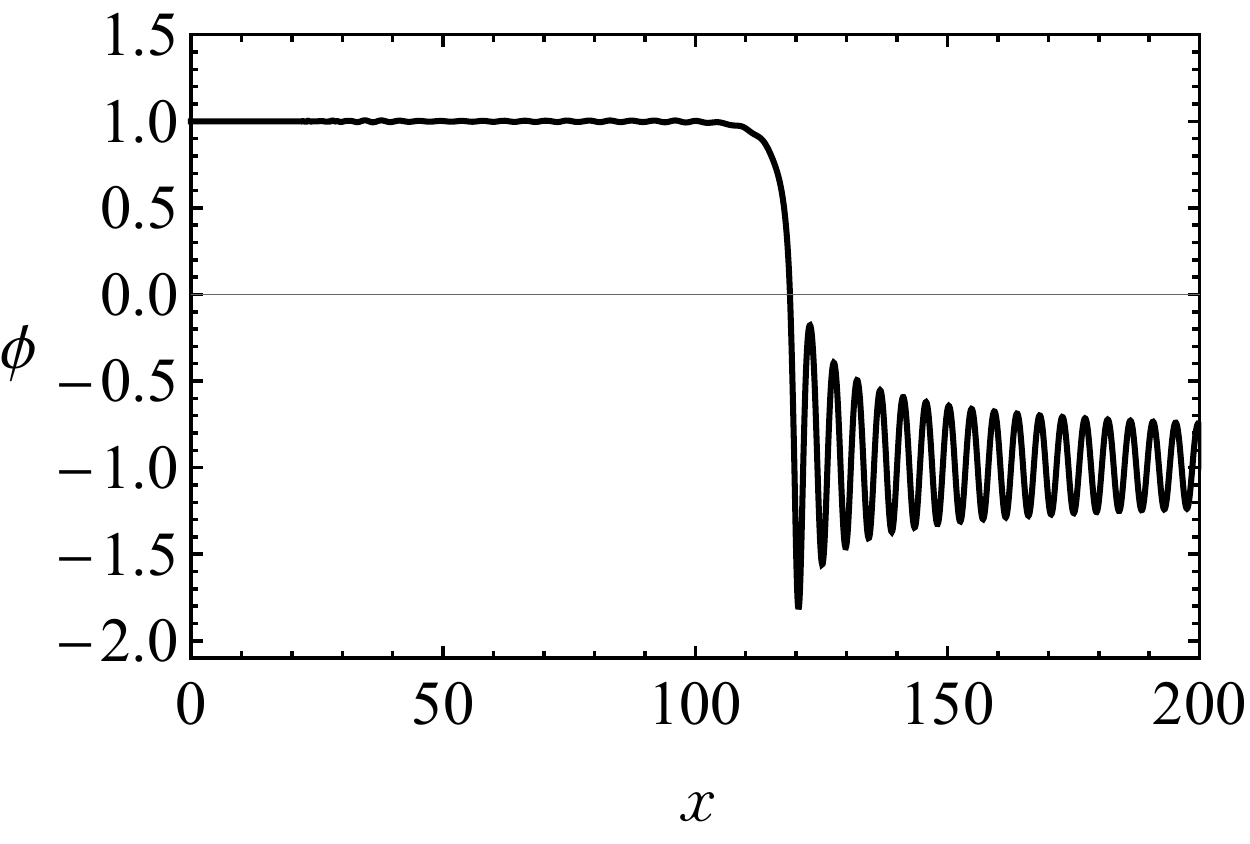}
}
\subfigure[\ $v_{\rm in}=0.5000$]{
\centering\includegraphics[width=
0.35\linewidth
]{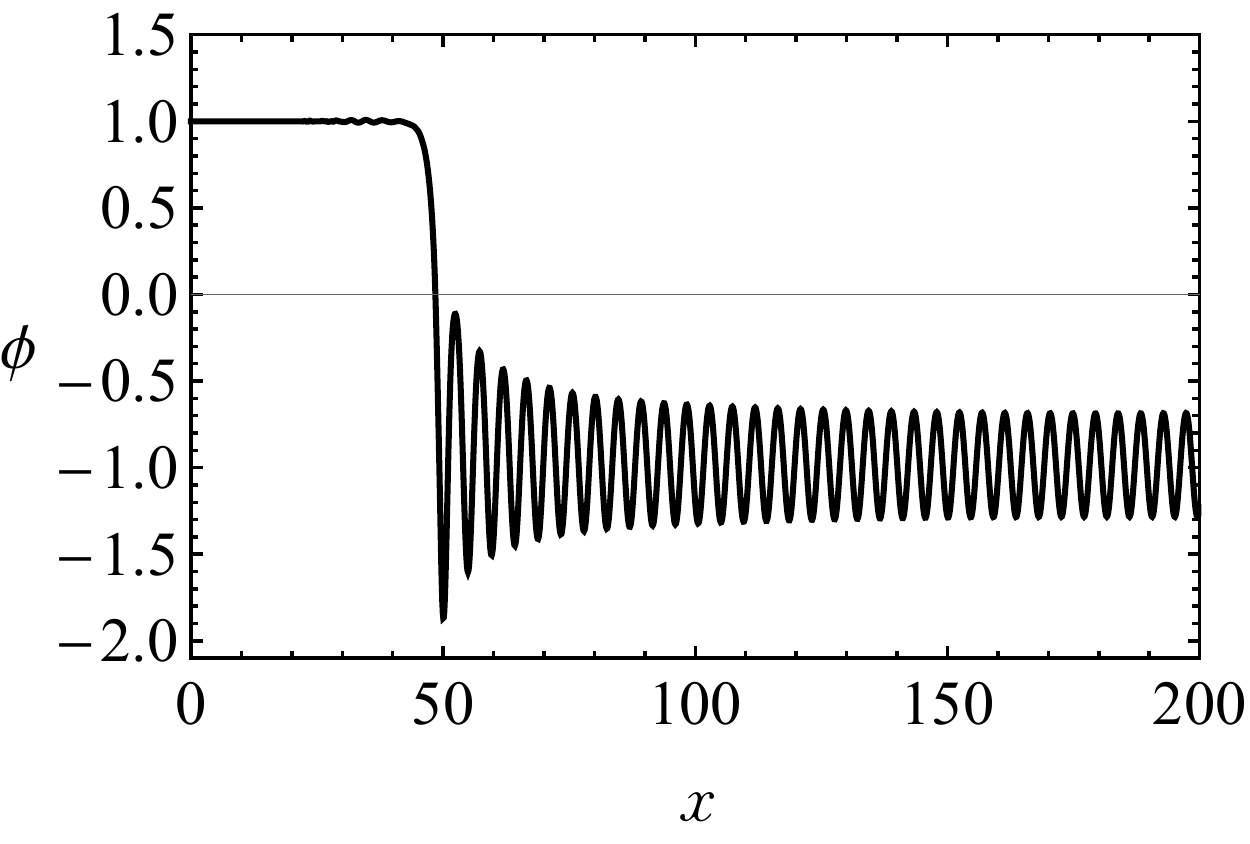}
}
\subfigure[\ $v_{\rm in}=0.7840$]{
\centering\includegraphics[width=
0.35\linewidth
]{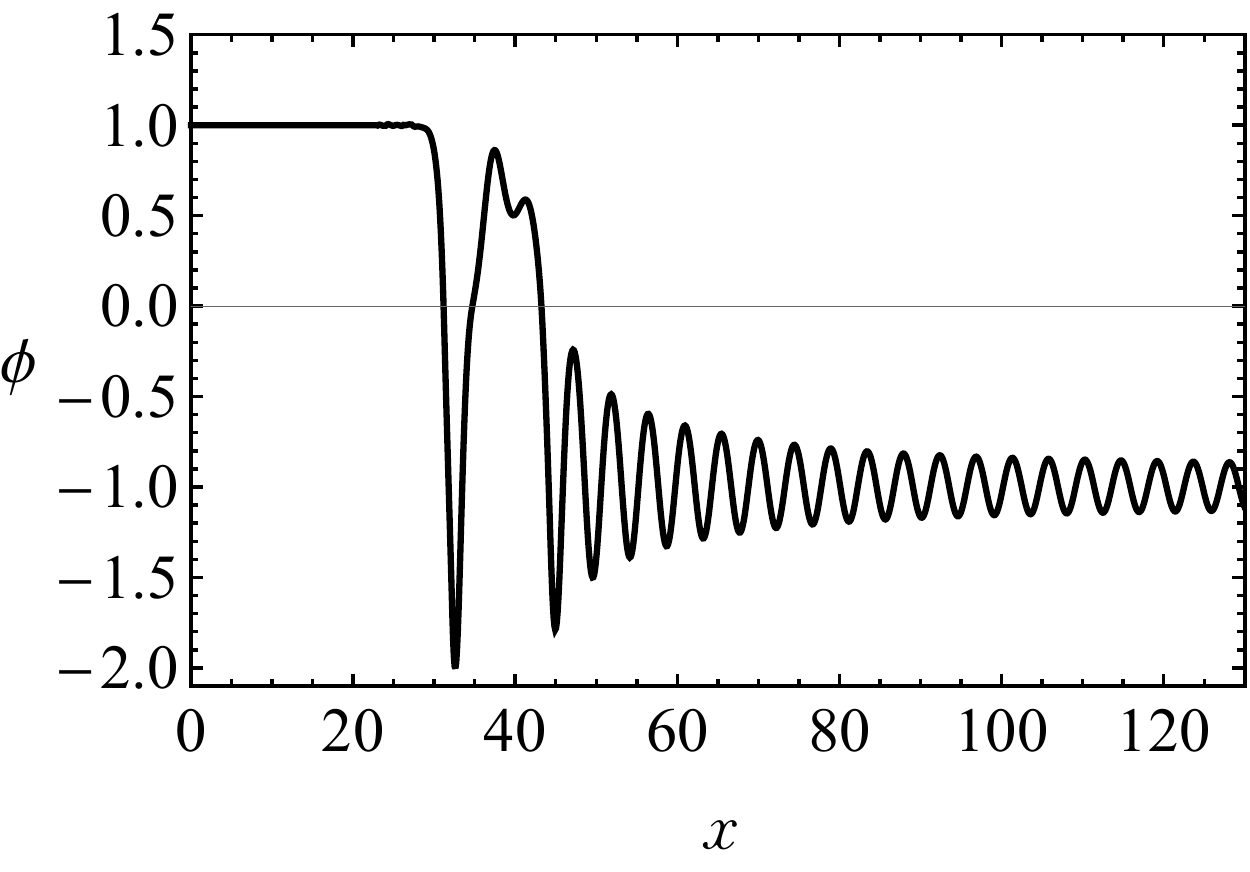}
}
\subfigure[\ $v_{\rm in}=0.7940$]{
\centering\includegraphics[width=
0.35\linewidth
]{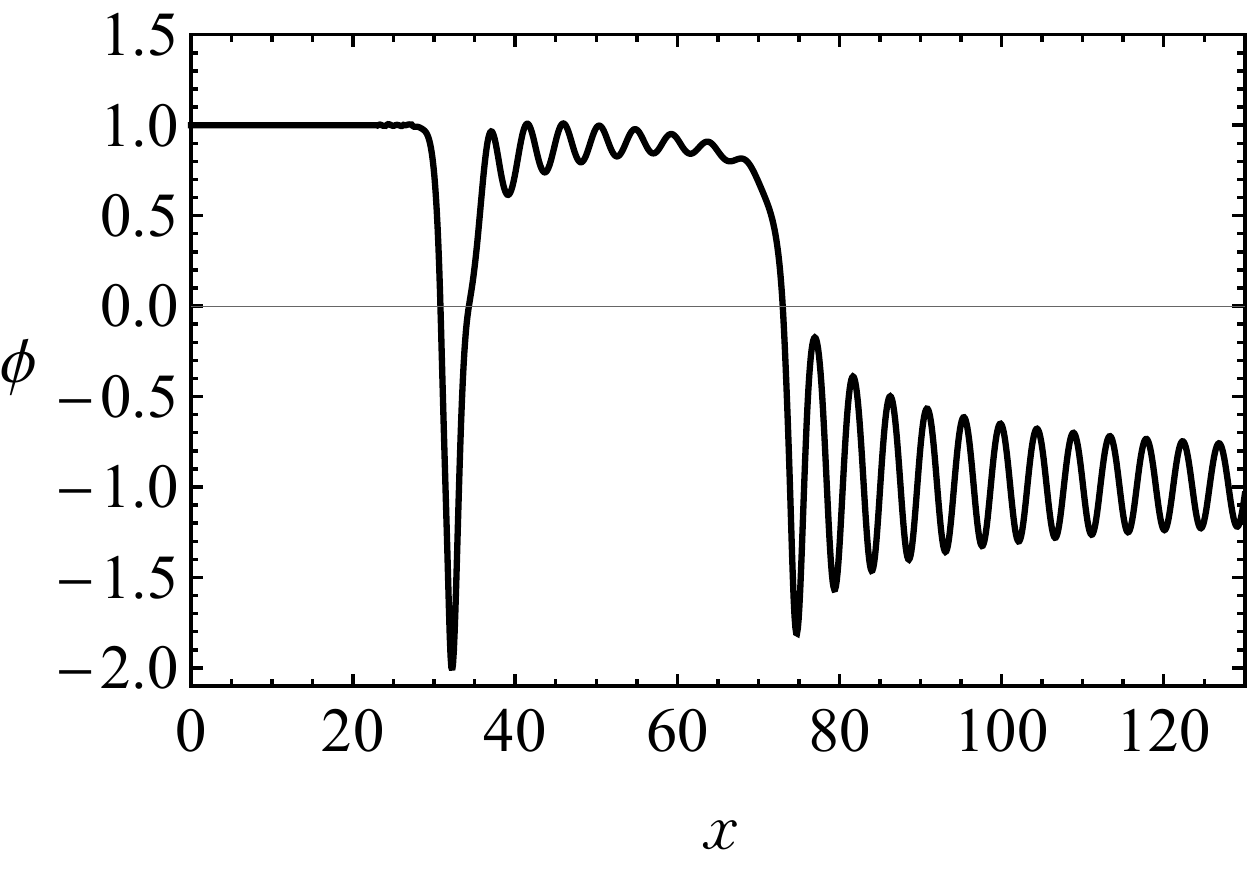}
}
\caption{Time dependence of the field at the collision point $x=0$ for the cases shown in Fig.~\ref{fig:bion}.}
\label{fig:time-x0-vbion}
\end{figure*}

We found a critical value of the initial velocity $v_\mathrm{cr}^{} \approx 0.79$, which separates two different collision regimes. At $v_\mathrm{in}^{} < v_\mathrm{cr}^{}$, the kinks annihilate radiating their energy away, in the form of small-amplitude waves, see Figs.~\ref{fig:bion} and \ref{fig:time-x0-vbion}.


\begin{figure*}[t!]
\subfigure[\ $v_{\rm in}=0.7950$]{
\centering\includegraphics[width=
0.35\linewidth
]{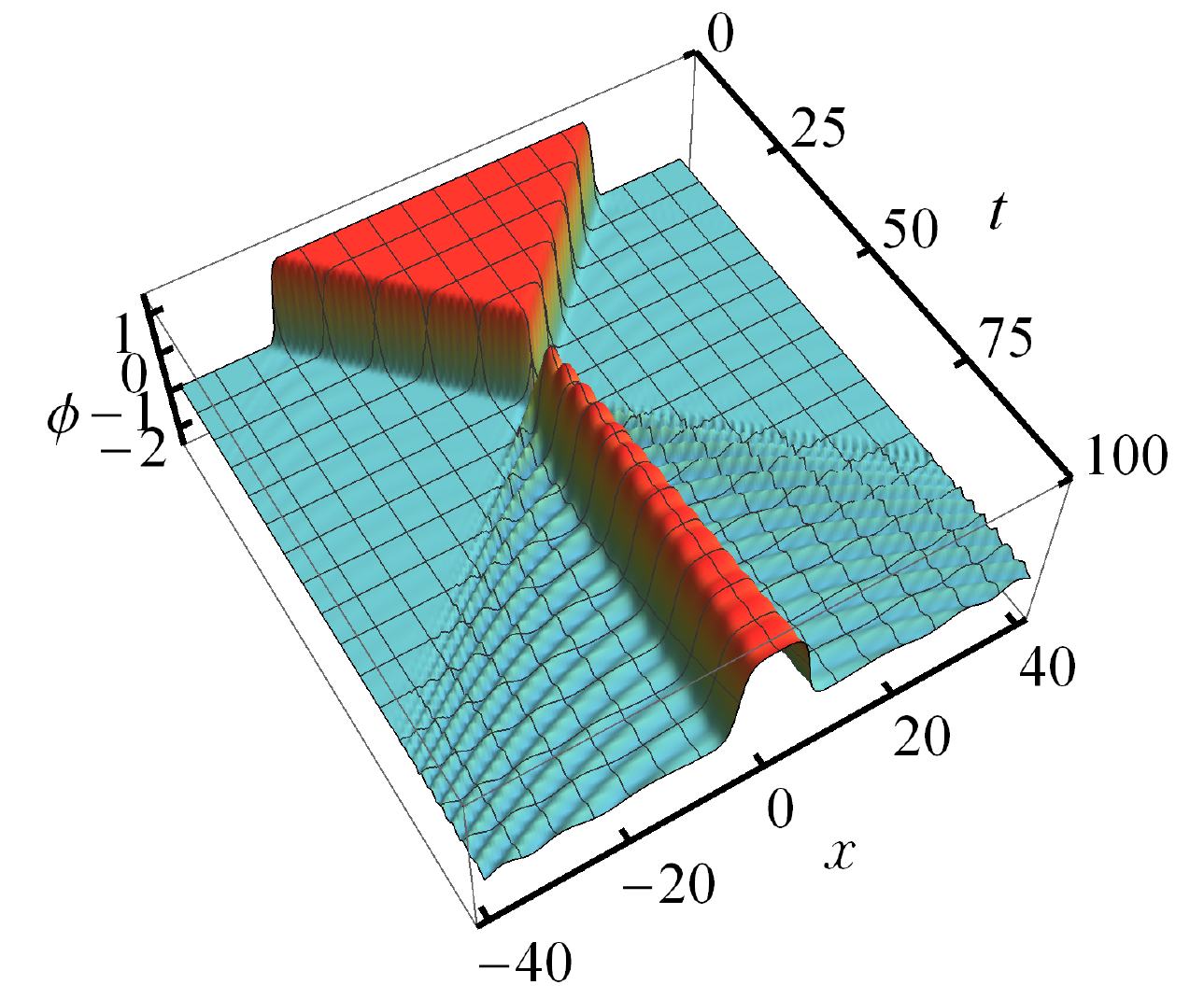}
}
\subfigure[\ $v_{\rm in}=0.7980$]{
\centering\includegraphics[width=
0.35\linewidth
]{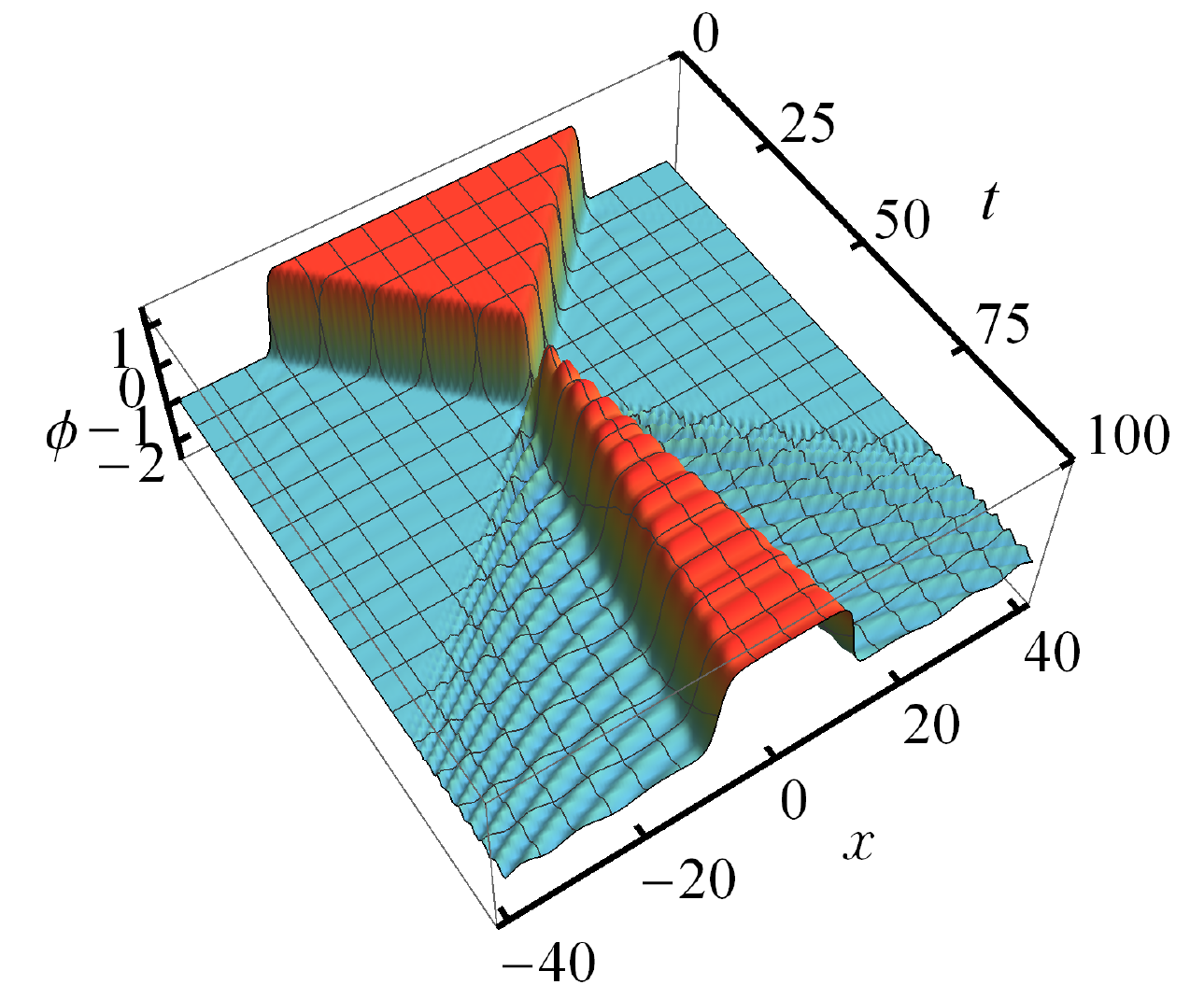}
}
\subfigure[\ $v_{\rm in}=0.8050$]{
\centering\includegraphics[width=
0.35\linewidth
]{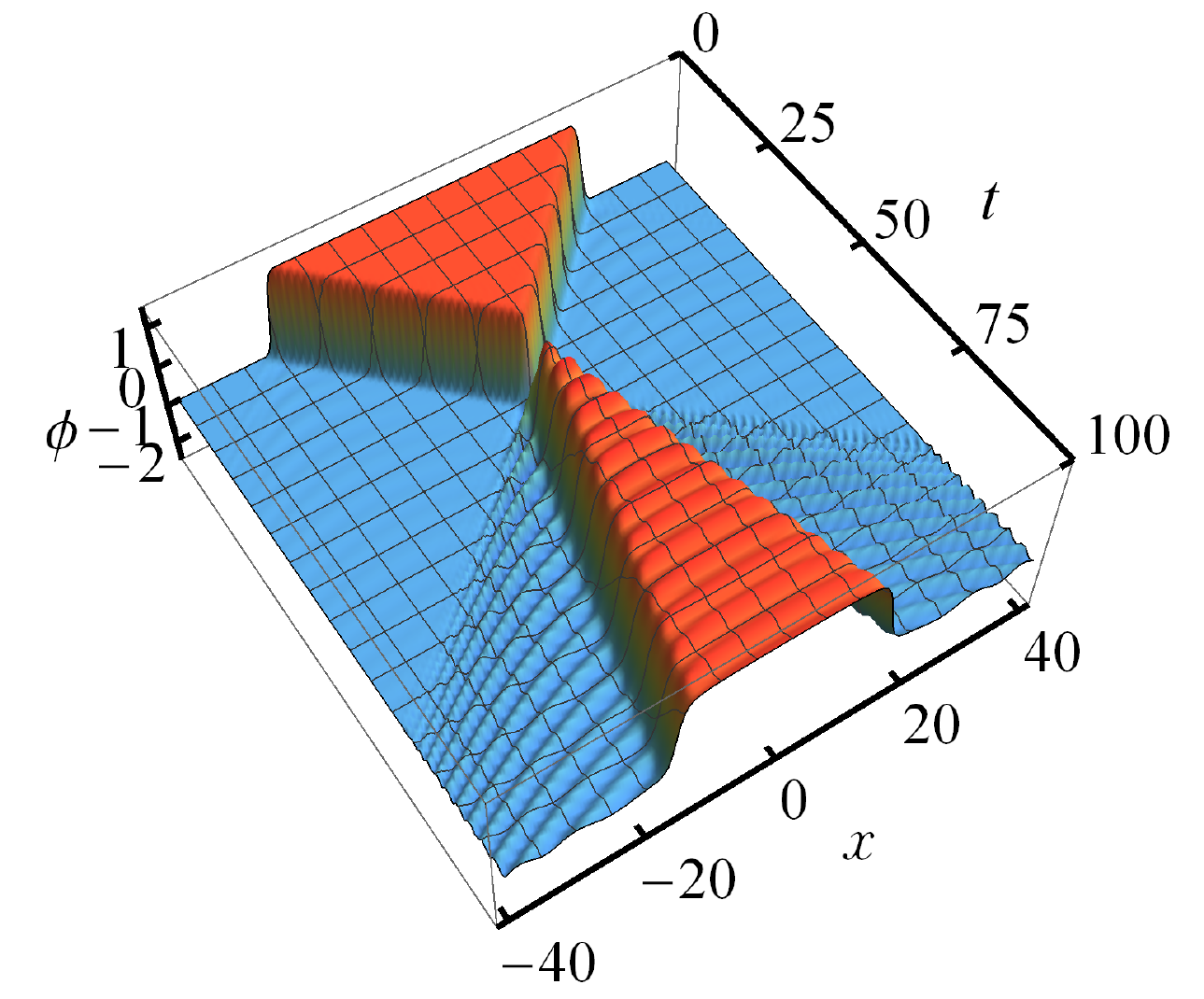}
}
\subfigure[\ $v_{\rm in}=0.8250$]{
\centering\includegraphics[width=
0.35\linewidth
]{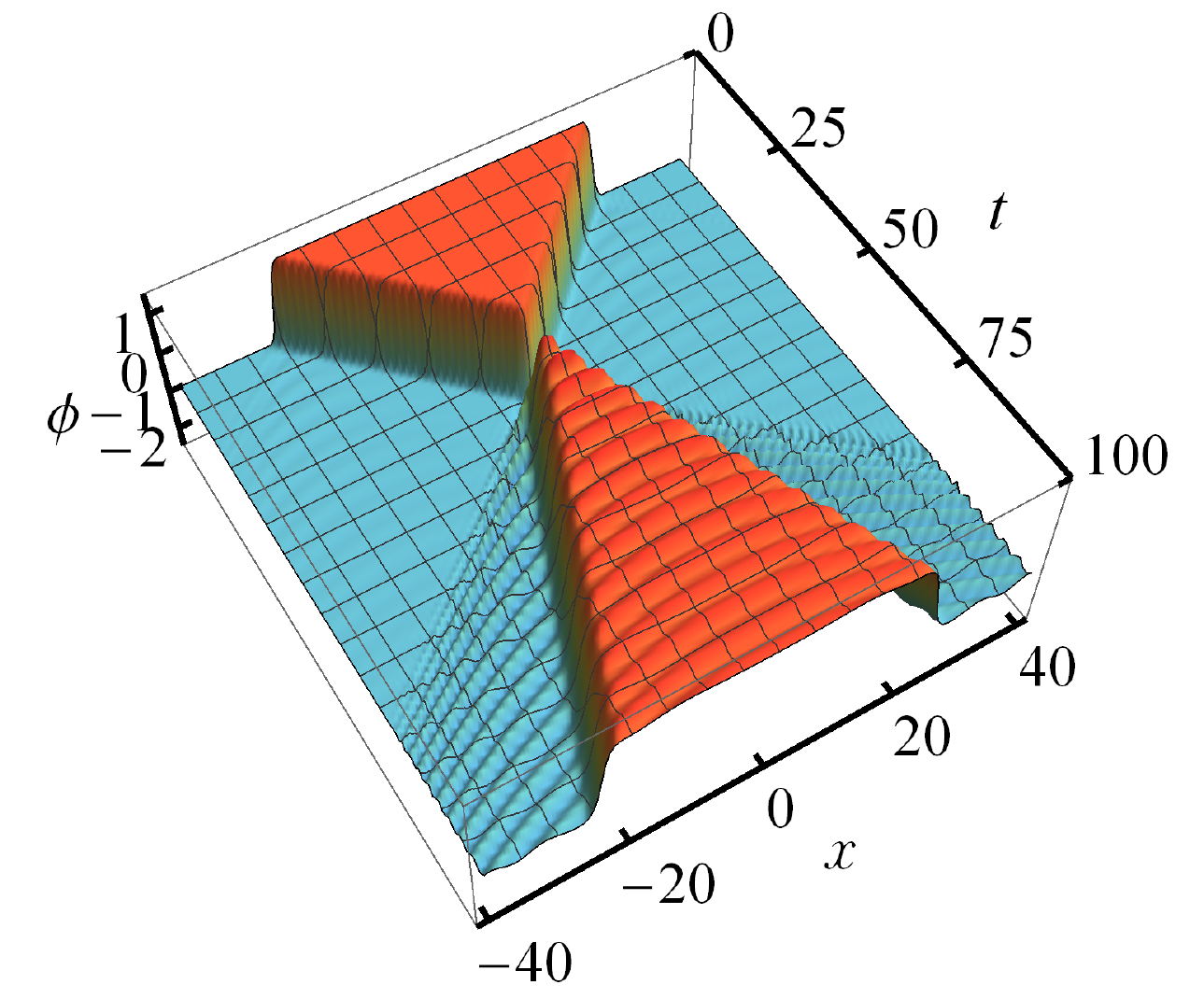}
}
\subfigure[\ $v_{\rm in}=0.9000$]{
\centering\includegraphics[width=
0.35\linewidth
]{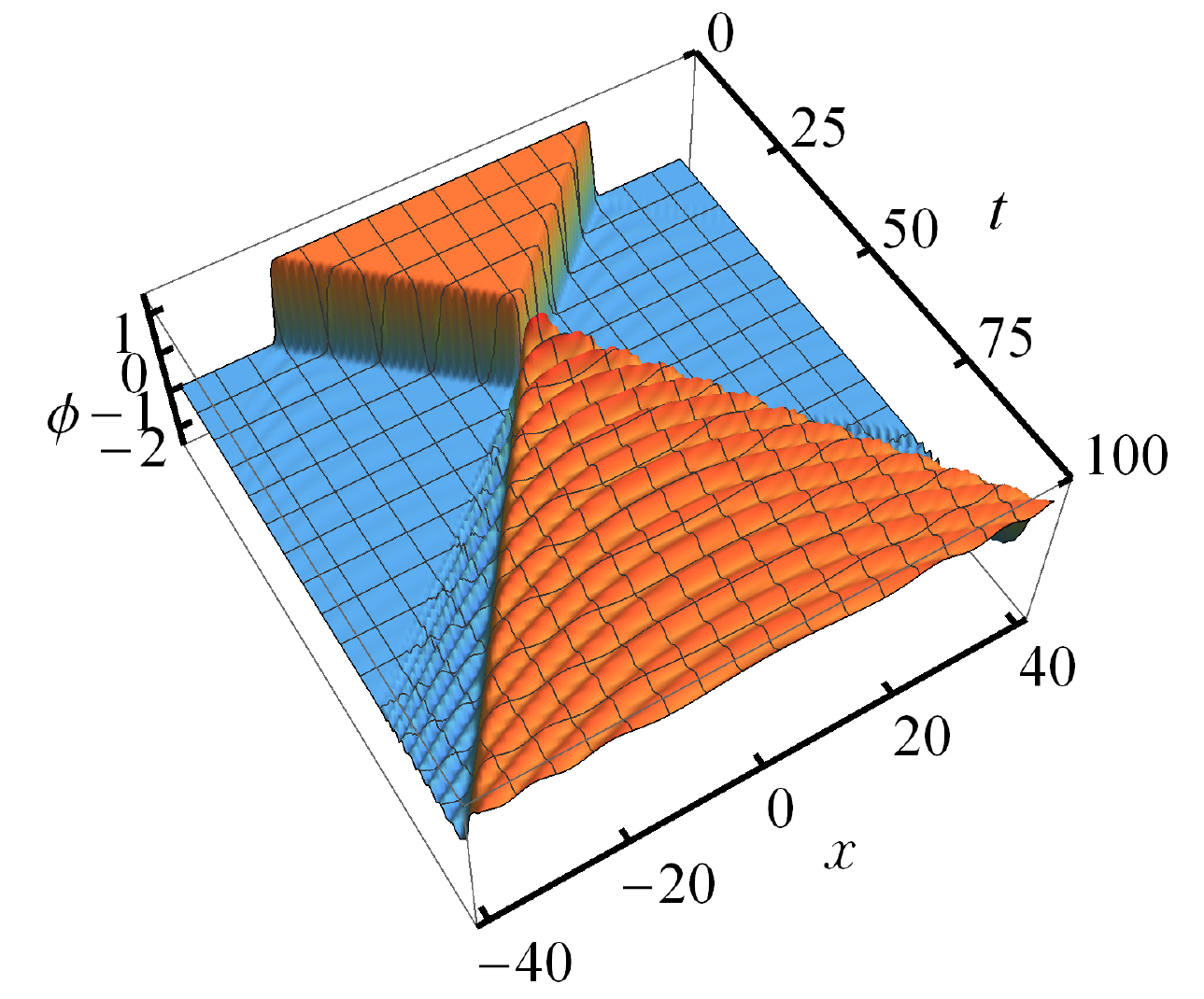}
}
\subfigure[\ $v_{\rm in}=0.9500$]{
\centering\includegraphics[width=
0.35\linewidth
]{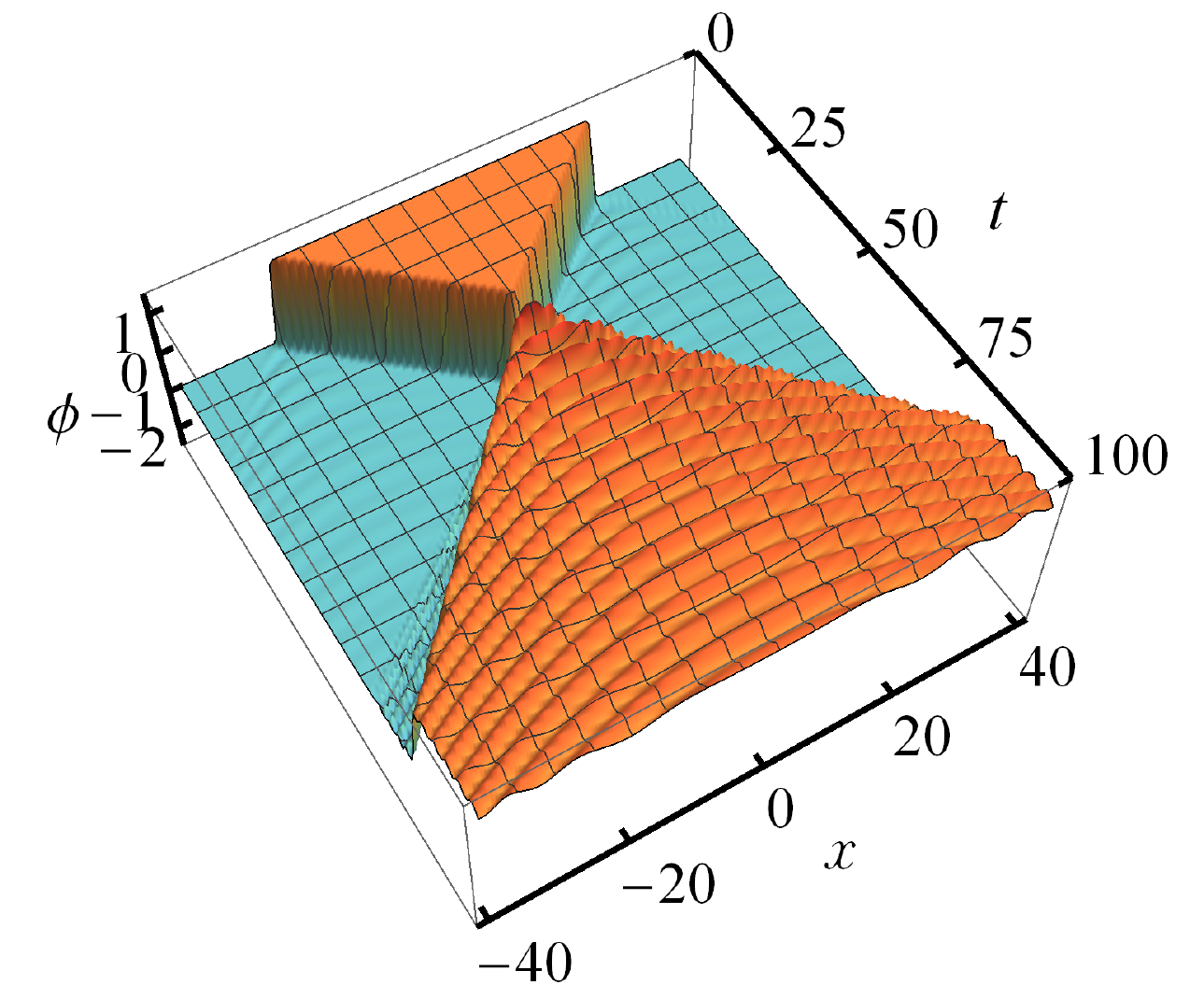}
}
\caption{Spacetime picture of the kink-antikink scattering at $v_{\rm in}^{}>v_{\rm cr}^{}$; $\ell=\sqrt{2}$, $\phie=1$.
}
\label{fig:time-coord}
\end{figure*}

\begin{figure*}[t!]
\subfigure[\ $v_{\rm in}=0.7950$]{
\centering\includegraphics[width=
0.35\linewidth
]{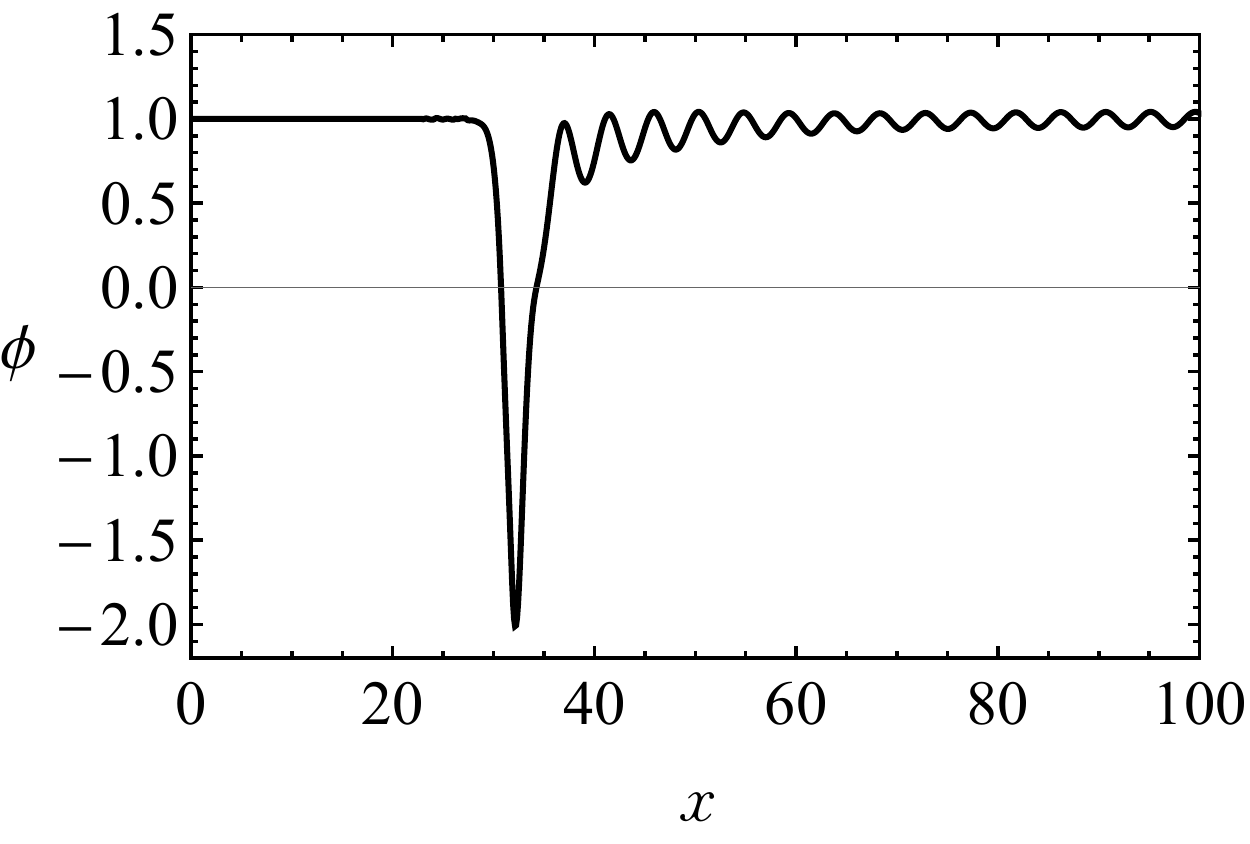}
}
\subfigure[\ $v_{\rm in}=0.7980$]{
\centering\includegraphics[width=
0.35\linewidth
]{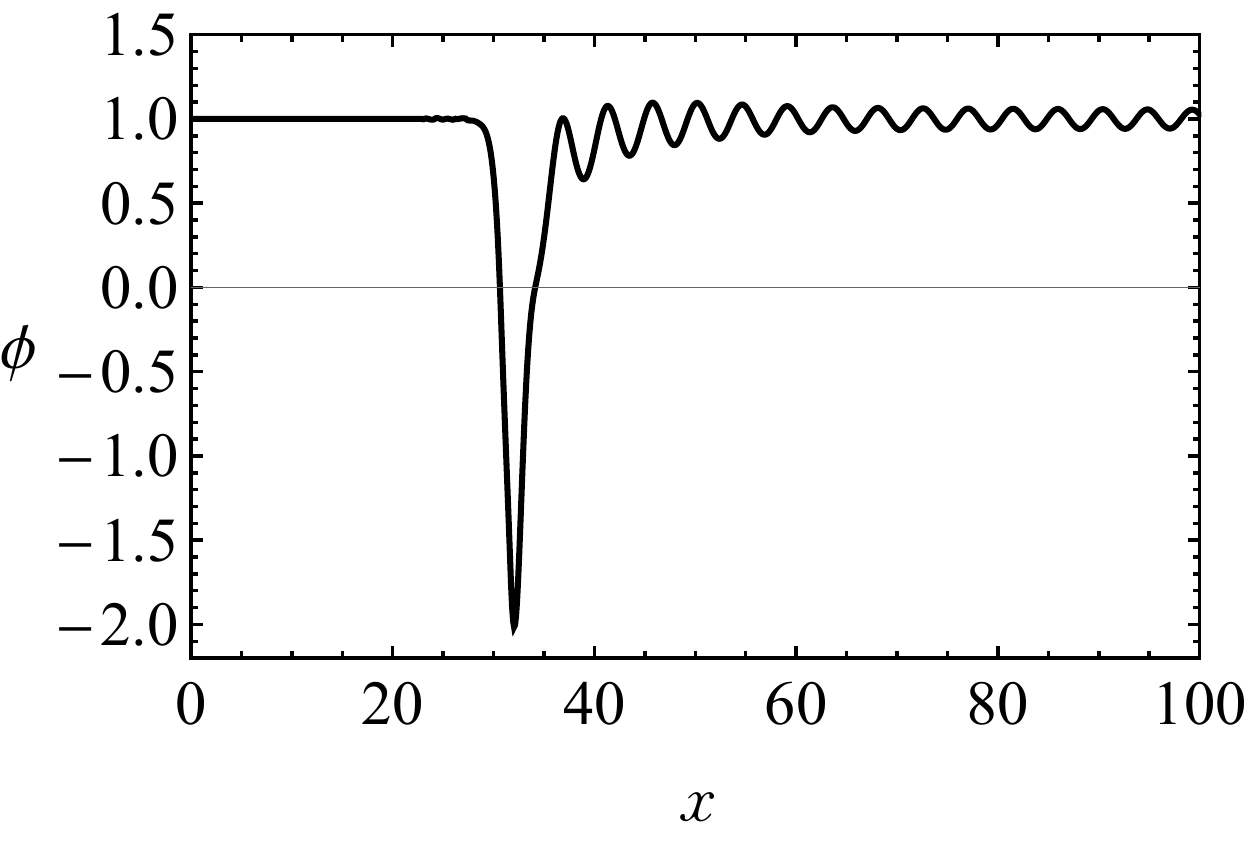}
}
\subfigure[\ $v_{\rm in}=0.8050$]{
\centering\includegraphics[width=
0.35\linewidth
]{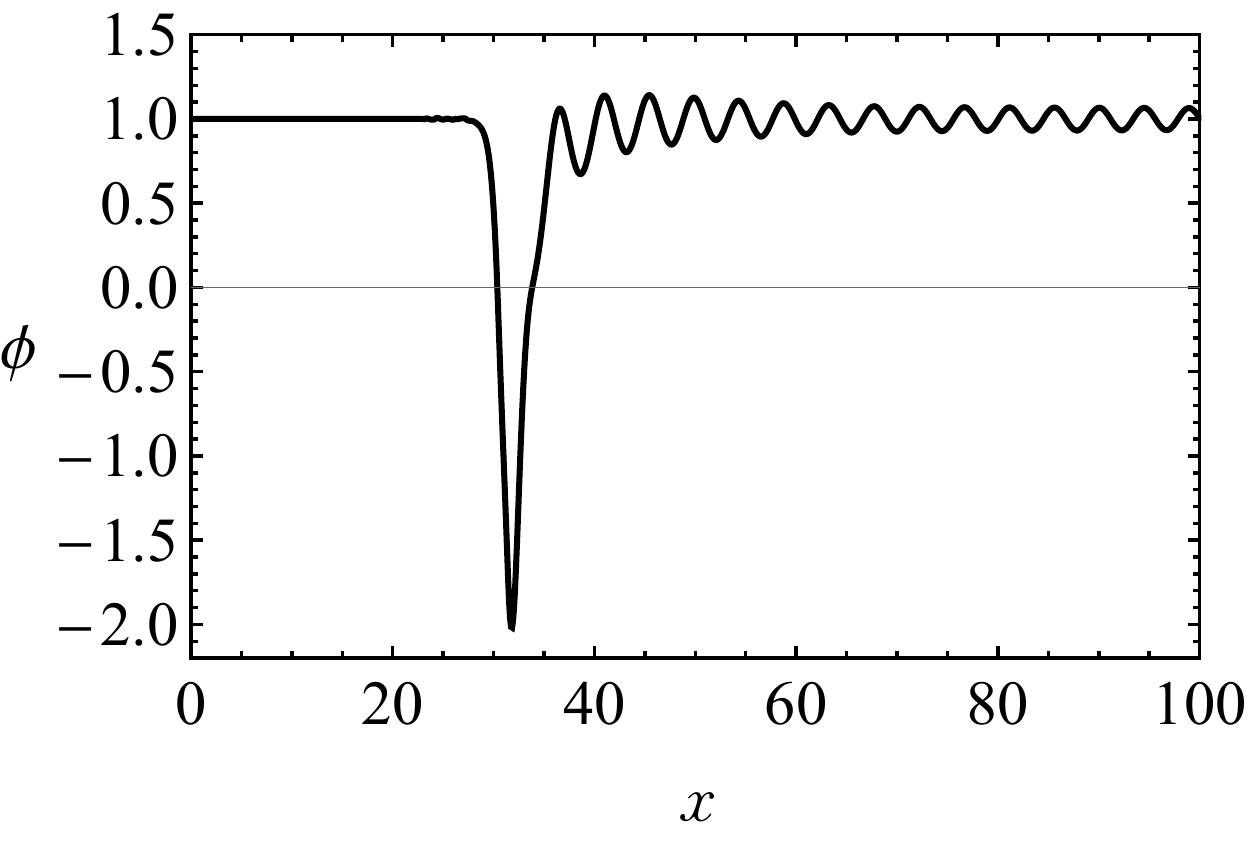}
}
\subfigure[\ $v_{\rm in}=0.8250$]{
\centering\includegraphics[width=
0.35\linewidth
]{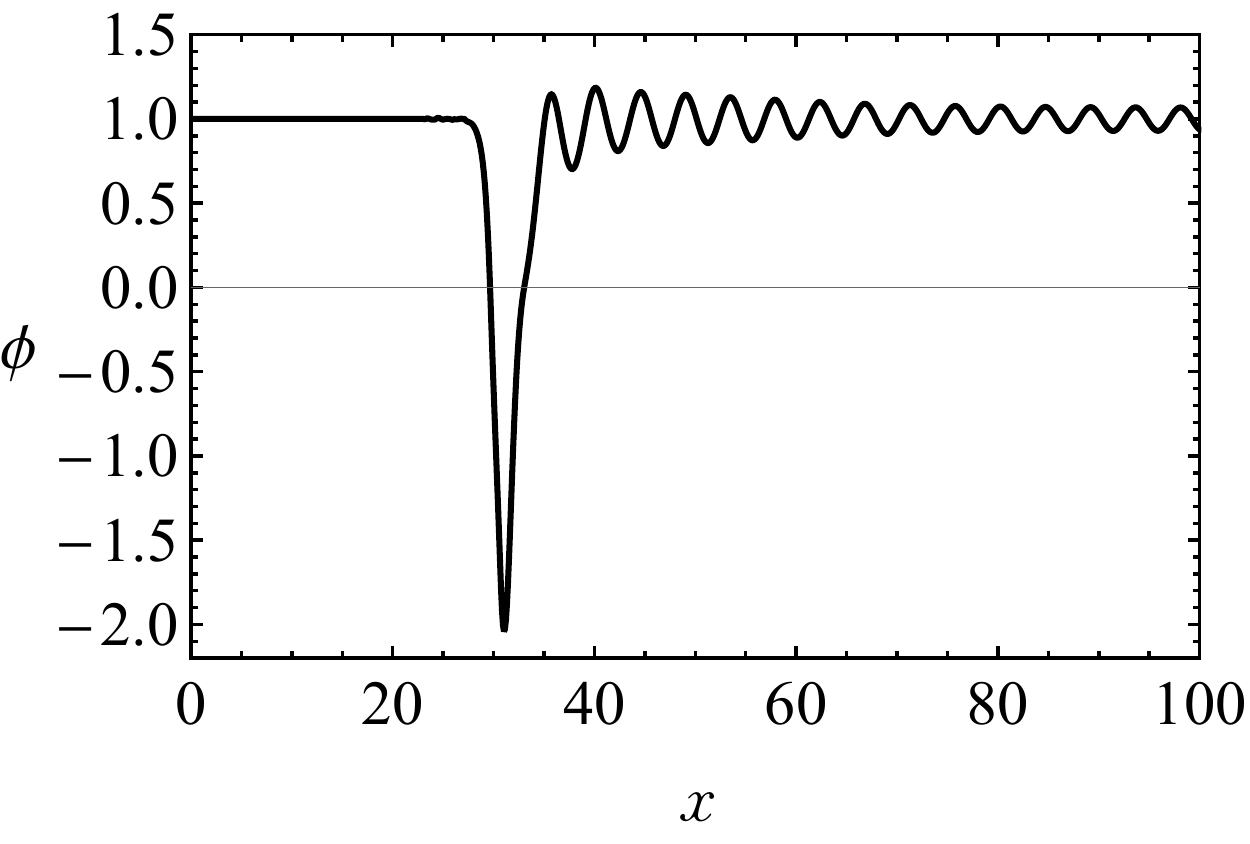}
}
\subfigure[\ $v_{\rm in}=0.9000$]{
\centering\includegraphics[width=
0.35\linewidth
]{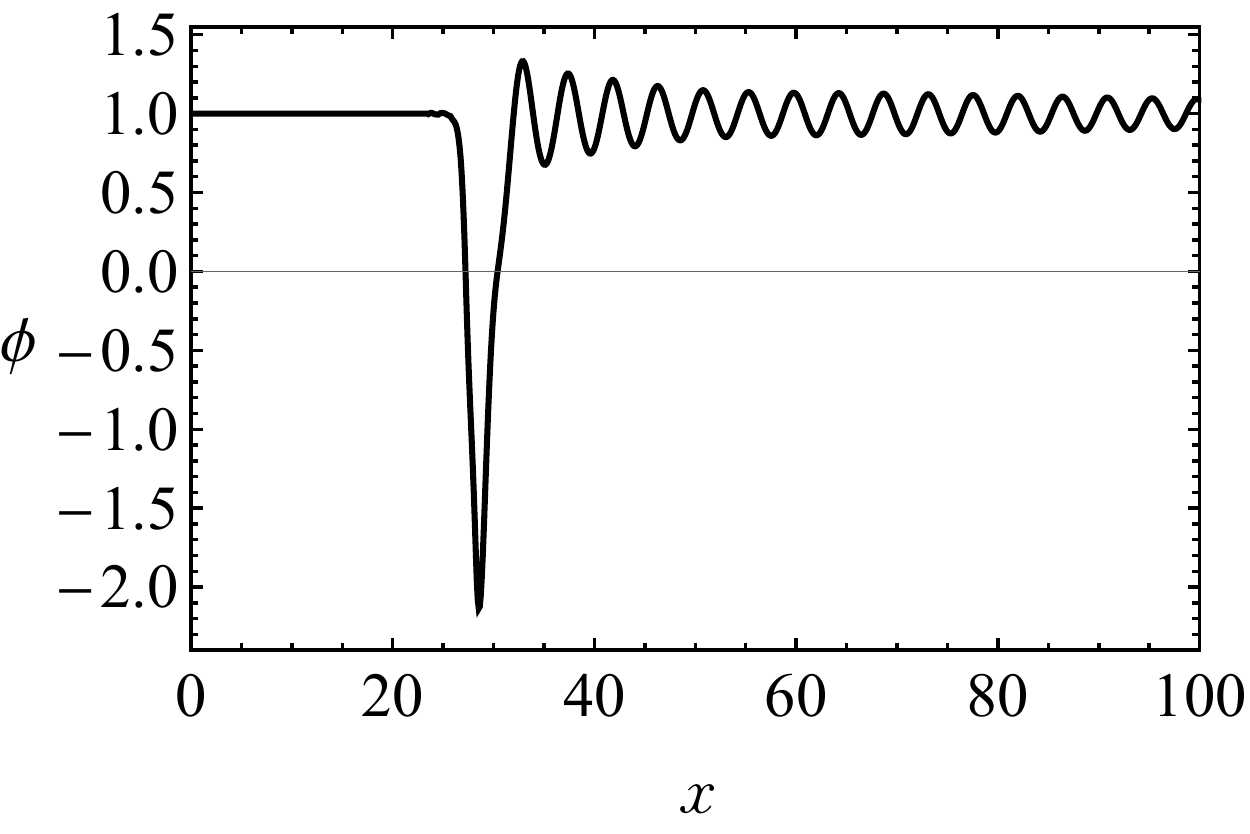}
}
\subfigure[\ $v_{\rm in}=0.9500$]{
\centering\includegraphics[width=
0.35\linewidth
]{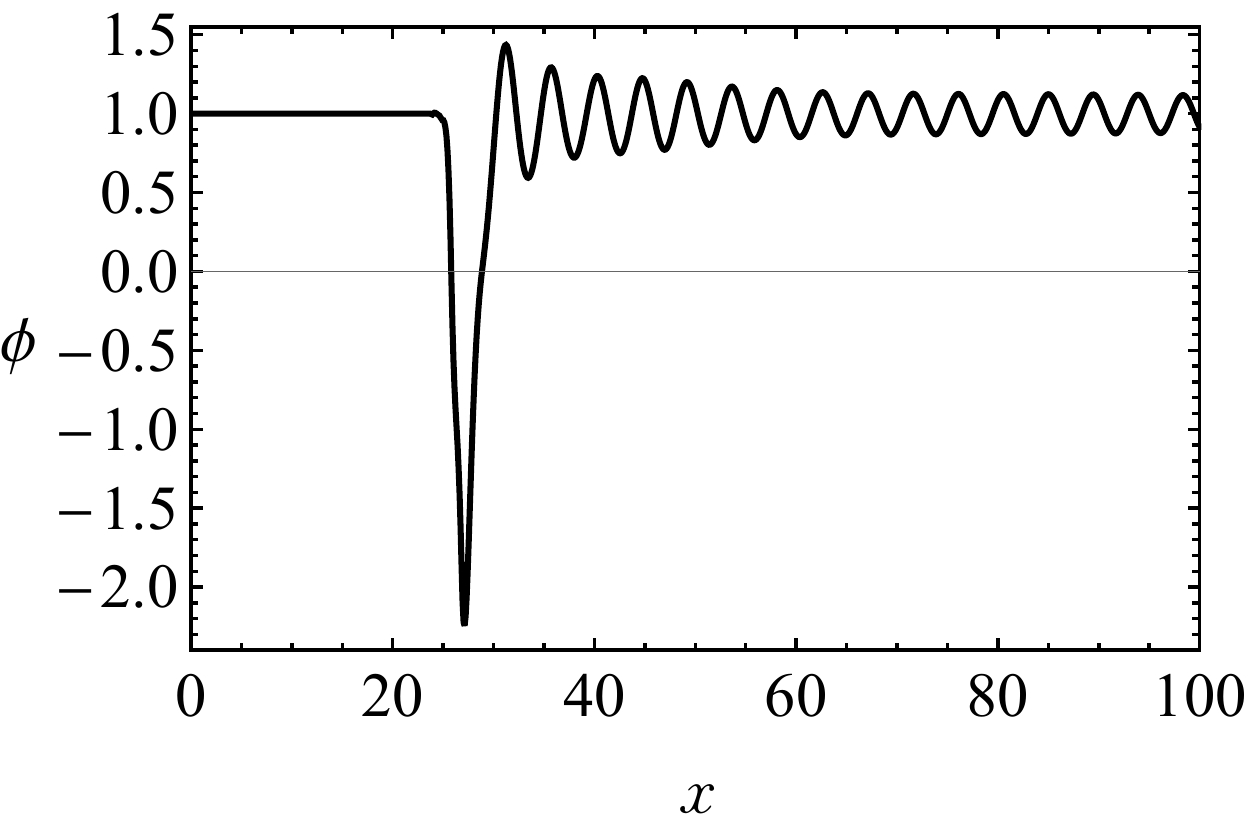}
}
\caption{Time dependence of the field at the collision point $x=0$ for the cases shown in Fig.~\ref{fig:time-coord}.}
\label{fig:time-x0-vescape}
\end{figure*}

On the other hand, at $v_\mathrm{in}^{} > v_\mathrm{cr}^{}$, we observe the inelastic reflection of kinks, see Figs.~\ref{fig:time-coord} and \ref{fig:time-x0-vescape}. 
The kink and antikink collide and escape from each other with final velocities $v_{\rm f}^{}<v_{\rm in}^{}$. 
We also observe that small-amplitude waves carry away some part of the kinks' kinetic energy.

We emphasize that in the range of initial velocities $v_{\rm in}^{}<v_{\rm cr}^{}$ we did not observe any resonance phenomena, such as escape windows, whose presence could indicate a resonant energy exchange between 
the translational and vibrational modes of kinks (see, e.g., Sec.~4 in \cite{Gani.JHEP.2015} or Sec.~IV in \cite{Belendryasova.CNSNS.2019} for a more detailed discussion of such phenomena in other models).
This result is consistent with the fact that the excitation spectrum of the kink contains only zero mode.

\scn{Remarks on logarithmic, quartic Higgs, and some other models}{s:higgs}

Because for the topological logarithmic model the field potential \eqref{e:potk} has a Mexican hat shape, this model is qualitatively similar to the quartic Higgs, or the $\phi^4$, model, which is widely used in the electroweak sector of the Standard Model and various cosmological models. In particular, the logarithmic model also has built-in spontaneous symmetry breaking, and various mass generation mechanisms  \cite{z11appb,dz12,z20ijmpa,z20un1,z21ltp}.

It is thus natural to compare kink's properties in these models.
To begin with, it is known that the $\phi^4$ kink has a non-singular smooth stability potential \eqref{eq:Schr_pot}, which is the well-known P\"oschl--Teller potential.
This potential can be shown to allow not only a ground state $\omega_0^{} = 0$, but also one excited state $\omega_1^{} > 0$, in the discrete spectrum. This excited state corresponds to the vibrational mode of the $\phi^4$ kink, which occurs in addition to the zero (translational) mode. As a result, energy transfer between these two modes can occur in various scattering processes involving the $\phi^4$ kinks \cite{Belova.UFN.1997}.

While the zero mode is connected with the translational invariance of the kink solution, the presence of the vibrational mode means possible existence of a long-living kink's excited state --- {\it wobbling kink}. Moreover, the presence of the vibrational mode in the kink's excitation spectrum leads to appearance of resonance phenomena in the kink-antikink collisions --- {\it escape windows} (sometimes called {\it bounce windows}) and {\it quasiresonances} (sometimes called {\it false windows}).

The escape windows are intervals of the initial velocity from the domain $v_\mathrm{in}^{}<v_\mathrm{cr}^{}$ ($v_\mathrm{cr}^{}\approx 0.2598$ for the $\phi^4$ kinks), withing which (that is at the initial velocities of the colliding kinks from these intervals) one observes escape of the kinks to spatial infinities after two or more impacts. This means that, despite the small initial velocity $v_\mathrm{in}^{} < v_\mathrm{cr}^{}$, the kinks are able to overcome mutual attraction. The cause of this is the resonant energy exchange between the two localized modes --- translational and vibrational. At the first impact, part of the kinks kinetic energy is transferred to the vibrational mode. As a result, the kinks are unable to overcome mutual attraction, they scatter at a finite distance and then collide again. If the frequency $\omega_1^{}$ and time between the first and the second impacts $T_{12}^{}$ satisfy a resonance condition, then a part of energy can be transferred from the vibrational mode back to the translational. It may be enough for the kinks to escape to spatial infinities. This scenario is called {\it two-bounce resonance}, while the interval of the initial velocities, in which this scenario is realized, is called {\it two-bounce escape window}, or simply {\it two-bounce window}.

Resonance transmission of energy from the vibrational mode to translational one can also happen in third and subsequent impacts. Such scenarios correspond to {\it three-bounce resonance}, and so on. It was found that the escape windows form a quasifractal structure --- near each two-bounce window there is a series of three-bounce windows, near each three-bounce window there is a series of four-bounce windows, and so forth. Notice that in some models the so-called quasiresonances --- maxima on the dependence of time between the second and the third impacts of the kinks on the initial velocity, $T_{23}^{}(v_\mathrm{in}^{})$ --- have been observed instead of the two-bounce windows, see, e.g., Figs.~1 and 2 in \cite{Gani.PRE.1999}.

On the contrary, the stability potential of the logarithmic kink, given by Eq.~\eqref{eq:Schr_pot_k}, allows only one state $\omega_0^{} = 0$. Therefore, no vibrational mode, hence no energy transfers between the modes, occur.

It is noteworthy that in the logarithmic model, the value of critical velocity turns out to be rather large: $v_{\rm cr}\approx 0.79$. By comparison, in the $\phi^4$ model $v_{\rm cr}\approx 0.26$, in the $\phi^6$ model $v_{\rm cr}\approx 0.05$ and $v_{\rm cr}\approx 0.29$ for different initial configurations. In the double sine-Gordon model, the critical velocity depends on the potential's parameter with the maximum value being $v_{\rm cr}\approx 0.24$, see Fig.~6 in \cite{Gani.EPJC.2018}.

It is usually assumed that the value of critical velocity shows how far is the model from being integrable. For example, in the integrable sine-Gordon model, critical velocity is formally equal to zero. If we look at the plot of critical velocity versus parameter $R$ in the double sine-Gordon model (Fig.~6 in \cite{Gani.EPJC.2018}), we can see that critical velocity vanishes both at $R=0$ and $R\to+\infty$. In both cases, the double sine-Gordon model transforms into the sine-Gordon model. At the same time, somewhere between $R=0$ and $R=+\infty$, the case farthest from integrability is realized, that corresponds to the global maximum on the plot of $v_{\rm cr}(R)$ dependence. Thus, we can say that the logarithmic model under consideration is in a sense far from integrable.

Another interesting fact is the very fast annihilation of the kink and antikink at $v_{\rm in}<v_{\rm cr}$, and significant loss of kinetic energy by kinks as a result of collision at $v_{\rm in}>v_{\rm cr}$. Both of these peculiarities are probably the manifestations of the same property of the logarithmic model. According to Fig.~\ref{fig:bion}, the decay of the kink-antikink bound state occurs almost without formation of a bion, especially when initial velocities are not too large, cf.\ Figs.~\ref{fig:bion}(a) and \ref{fig:bion}(b). Furthermore, significant loss of kinetic energy of kinks at $v_{\rm in}>v_{\rm cr}$ is clear from Fig.~\ref{fig:time-coord}. Final velocities of the kink and antikink are much less than initial velocities thereof, especially when the latter are not too large, cf.\ Figs.~\ref{fig:time-coord}(a)--\ref{fig:time-coord}(d).

Finally, in the logarithmic model, unlike other models mentioned above, the kink's stability potential is not bounded from below, it has singularity at the center of the kink. However, this singularity does not lead to appearance of modes with negative $\omega^2$.

\scn{Conclusion}{s:con}

In this paper, we studied the properties of the kink solution in a relativistic field-theoretic model with logarithmic potential.
The static kink occurs as a solution of the first order ordinary differential equation, which was solved numerically.

It turns out that the kink's excitation spectrum has only zero (translational) mode. This probably means that no resonance phenomena can be found in the kink-antikink collisions. However, an absence of vibrational mode(s) does not guarantee an absence of resonance phenomena. Therefore, we performed numerical simulations of the kink-antikink collisions in a wide range of their initial velocities. We found the critical value of the initial velocity, $v_\mathrm{cr}^{}\approx 0.79$, which separates two different collision regimes.

At $v_{\rm in}^{}<v_{\rm cr}^{}$, we observed the capture of kinks and the formation of a bion (a kink-antikink bound state). Subsequently, this bion promptly decays by radiating energy in the form of small-amplitude waves. In other words, the kink and antikink annihilate for this range of initial velocities. It is noteworthy that annihilation occurs rather quickly, hence the bion state occurs only at large initial velocities and has a predominantly short lifetime. Moreover, in our numerical simulations, we did not observe any resonance phenomena, such as escape windows or quasiresonances. This fact is in agreement with the absence of vibrational modes.

At $v_\mathrm{in}^{}>v_\mathrm{cr}^{}$, kinks bounce off each other and escape to infinity, while radiating part of their energy in the form of small-amplitude waves. It is noteworthy that the loss of kinetic energy in such collisions is quite large.

A remarkable feature of the considered model is large critical velocity compared to other non-integrable models with polynomial and non-polynomial potentials. We conjecture that the high critical velocity, fast annihilation of the kink and antikink at $v_\mathrm{in}^{}<v_\mathrm{cr}^{}$, and the high inelasticity of collisions at $v_\mathrm{in}^{}>v_\mathrm{cr}^{}$ are all manifestations of the same feature of the logarithmic model. A detailed study of this conjecture could be a subject of future work.

\begin{acknowledgments}

This work was supported by the Russian Foundation for Basic Research under Grant No.\ 19-02-00971. The work of the MEPhI group was also supported by MEPhI within the Program ``Priority-2030''. Numerical simulations were performed using resources of NRNU MEPhI high-performance computing center. K.G.Z.'s research is supported by the Department of Higher Education and Training of South Africa and in part by the National Research Foundation of South Africa (Grants Nos.\ 95965, 131604 and 132202). Proofreading of the manuscript by P.~Stannard is greatly appreciated.

\end{acknowledgments}

\end{document}